\documentclass[amsfonts,amsmath,prd,preprint,nofootinbib,superscriptaddress]{revtex4}
\usepackage[dvipdfmx]{graphicx}
\usepackage{bm,latexsym,amsmath,amssymb,amsfonts} 
\DeclareMathAlphabet{\mathpzc}{OT1}{pzc}{m}{it}
\usepackage[normalem]{ulem}
\usepackage[breaklinks=true]{hyperref}
\usepackage{xcolor}
\usepackage{wasysym}
\usepackage[mathscr]{eucal}
\usepackage{multirow}
\usepackage{enumitem}
\usepackage{tabularray}
\usepackage{tablefootnote}
\usepackage{longtable}
\usepackage{float}
\usepackage{booktabs}
\usepackage{lipsum}

\begin{document}
\title{Anisotropic gravastar as horizonless regular black hole spacetime and its images illuminated by thin accretion disk}

\author{M.~F.~Fauzi}
\email{muhammad.fahmi31@ui.ac.id}
\affiliation{Departemen Fisika, FMIPA, Universitas Indonesia, Depok 16424, Indonesia}

\author{H.~S.~Ramadhan}
\email{hramad@sci.ui.ac.id}
\affiliation{Departemen Fisika, FMIPA, Universitas Indonesia, Depok 16424, Indonesia}

\author{A.~Sulaksono}
\thanks{Corresponding author}
\email{anto.sulaksono@sci.ui.ac.id}
\affiliation{Departemen Fisika, FMIPA, Universitas Indonesia, Depok 16424, Indonesia}
 
\begin{abstract}
A connection between regular black holes and horizonless ultracompact objects was proposed in~\cite{Carballo-Rubio:2022nuj}. In this paper, we construct a model of a horizonless compact object, specifically an anisotropic gravastar with continuous pressure, that corresponds to regular black hole spacetime in the appropriate limit. The construction begins by modeling an equation of state that satisfies the anisotropic gravastar conditions and transitions to the de Sitter ($p=-\epsilon$) upon horizon formation. The spacetime structure is similar to the {\it Quantum Horizonless Compact Object} (QHCO) described in~\cite{Chen:2024ibc}. Within this model, we also generate images of the corresponding objects surrounded by a thin accretion disk. The resulting images reveal that assuming that the emitting matter exists only outside the object, the inner light ring structure closely resembles that of the horizonless configuration of a regular black hole and the QHCO, yet it exhibits a distinct light ring structure compared to the thin-shell gravastar model. However, the opposite occurs when emitting matter is taken into account inside the object.
\end{abstract}

\maketitle
%------------------------------------------------
\section{Introduction}

Einstein’s theory of general relativity predicts the existence of two irregular points where the corresponding metric blows up: the horizons and the singularities, leading to the definition of a black hole (BH). BH can be ``seen" in numerous observational circumstances, including the image of a supermassive BH at the center of the Milky Way. For extensive discussions on this topic please refer to Ref.~\cite{DuttaRoy:2022ytr} and the references therein. According to the {\it Weak Cosmic Censorship Conjecture} (WCCC), the singularities are always hidden behind horizons, preventing them from being observed directly by a distant observer~\cite{Penrose:1969pc}. Consequently, certain parameters of black holes, such as those in Kerr and Reissner-Nordstrom black holes, are not allowed to exceed the extremal limit. However, several thought-experiments show that it is possible to destroy the event horizon of several models of black holes by sending a test particle, thus exposing the singularity \cite{Hubeny:1998ga,Jacobson:2009kt,Izumi:2024rge,Jiang:2023gpx,Siahaan:2022fht,Siahaan:2015ljs}.

Several approaches have been attempted to resolve the singularity and horizon problems. A regular BH (RBH) without singularity was first proposed by Bardeen \cite{Bardeen}. The solution introduces a new parameter that regulates the BH. In general, the classical source of regularity can be produced in two ways: either by introducing a new physics scale (e.g., in Refs.~\cite{Hayward:2005gi,Cadoni:2022vsn,Dymnikova:1992ux}) or by incorporating an appropriate non-linear electrodynamics source which gives a regular metric solution (e.g. in Refs.~\cite{Ayon-Beato:2000mjt,Ayon-Beato:1998hmi,Bronnikov:2000vy,Burinskii:2002pz,Bronnikov:2021uta}). RBH also has an extremal limit where one might obtain a horizonless spacetime solution that results in a ``star" configuration rather than a naked singularity~\cite{Carballo-Rubio:2022nuj}. A recent review on RBH can be seen in Refs~\cite{Lan:2023cvz,Bambi:2023try}.

As an attempt to eliminate the need for horizons and singularities, Mazur and Motolla proposed {\it gravastar} solution as a BH mimicker~\cite{Mazur:2001fv}. In their formulation, the horizon is replaced by a thin shell of positive pressure matter while the interior of the metric is filled with de Sitter vacuum with negative pressure. The gravastar is known to fulfill the criteria of physical relevance, particularly in terms of its stability and properties that deviate only slightly from those of BHs, making it a promising alternative to BHs (see Ref.~\cite{Rosa:2024bqv} and the references therein). Numerous gravastar models have been studied in the literature \cite{Adler:2022fqu,Sanjay:2024mkg,Ghosh:2023wps,Jampolski:2023xwh,Moti:2021vck,Bhatti:2021xqi,Beltracchi:2021lez,Horvat:2008ch}, including those that feature no discontinuities in radial pressure dubbed as `\textit{anisotropic gravastar}' \cite{Cattoen:2005he,DeBenedictis:2005vp}. For a review of recent developments and status on gravastar, see Ref.~\cite{Ray:2020yyk}.

Horizonful and horizonless spacetimes have distinct observational signatures, notably in their optical images. It has been shown that by neglecting the photon interaction with matter on the interior, horizonless spacetime potentially produces several inner light rings \cite{Chen:2024ibc,Rosa:2023qcv,Rosa:2023hfm,Rosa:2024eva,Guerrero:2022msp,Eichhorn:2022fcl}, including in gravastar spacetime \cite{Rosa:2024bqv}, due to the deflection of light near the source of gravity. In contrast, the existence of horizons leads to a dark patch on the optical images larger than the horizon's radius, referred as the `shadow' (for a review, see Ref.~\cite{Cunha:2018acu}). It has also been demonstrated that horizonless spacetimes can exhibit a dark patch region due to gravitational redshift effects, though it is smaller compared to that of  spacetimes with horizon \cite{Chen:2024ibc,Rosa:2024bqv,Rosa:2023qcv,Rosa:2023hfm}. With the remarkable development of the Event Horizon Telescope (EHT) in the past decades, it has become possible to observe the distinction in the optical images with more advanced instruments, such as the next-generation EHT (ngEHT)~\cite{Eichhorn:2022fcl}.

To the best of our knowledge, horizonless solutions of regular black holes (RBHs) have not been extensively studied in the literature. While there are many proposed horizonless compact objects (for a review, see Ref.~\cite{Cardoso:2019rvt}), none of them explain their relation to the horizonless configuration of regular black holes. However, Carballo-Rubio et al. \cite{Carballo-Rubio:2022nuj} suggest that it might be possible to establish a connection between regular black holes and horizonless stars. Interestingly, the transverse pressure profile of the horizonless star configuration\textemdash considering it is sourced by anisotropic perfect fluid\textemdash matches the requirements of the anisotropic gravastar solution with continuous pressure.

The RBH's star solution, however, has a key issue: it lacks a clearly defined radius or boundary separating the interior from the exterior regions. In contrast, the boundary in the horizon solution is defined by the horizon's radius itself. Furthermore, with a repulsive de Sitter core, the matter would eventually accumulate around the core, forming a region of positive pressure \cite{Cattoen:2005he}. This scenario might not occur in a configuration with a horizon, as the horizon itself would immediately absorb matter. Therefore, a complete model of RBH spacetime that can describe both the horizon and star solutions is desirable. A promising candidate for such a star solution is the previously mentioned gravastar, especially the anisotropic gravastar model proposed by Cattoen, Faber, and Visser~\cite{Cattoen:2005he} and later by DeBenedictis et al. ~\cite{DeBenedictis:2005vp}. This possibility has yet to be explored, motivating us to construct an RBH model that transitions into an anisotropic gravastar at the extremal limit. In this paper, we shall also calculate its shadow images involving an optically and geometrically thin accretion disk and consider the effects of accretion flow on the inclined observation to obtain a more realistic scenario.

The discussion in this paper is arranged as follows. Sec.~\ref{sec. Review RBH grav} briefly reviews the regular black hole and gravastar solution. In Sec.~\ref{sec. Transition Construct}, we construct a regular black hole model that transitions into an anisotropic gravastar solution in the extremal limit. In Sec.~\ref{sec. Geodesic and shadow image} we calculate the shadow of our corresponding model, including the construction of the accretion disk model and the photon trajectory around the object, and compare them with the horizonless RBH spacetime. Finally, in Sec. \ref{sec. conclusion}, we summarize and discuss the results of our findings.

%=======================================================================
\section{Regular black holes and gravastar: a brief review}
\label{sec. Review RBH grav}

\subsection{Einstein equations  for anisotropic objects}

We consider a static and spherically symmetric spacetime with line element in form of
\begin{equation}
	ds^2 = -e^{\nu}dt^2 + e^{\lambda}dr^2 + r^2d\Omega^2.
	\label{eq. SSS ansatz enu}
\end{equation}
The Einstein field equation reads
\begin{equation}
	G^{\mu}_{\nu} = 8\pi T^{\mu}_{\nu}.
	\label{eq. EFE}
\end{equation}
For a perfect fluid with anisotropic pressure, the energy-momentum tensor satisfy the relation
\begin{equation}
	T^{\mu}_{\nu} = \left(\epsilon + p_t\right) u^{\mu}u_{\nu} + p_t \delta^{\mu}_{\nu} + (p-p_t) s^{\mu}s_{\nu},
\end{equation}
where $\epsilon$, $p$, and $p_t$ are the energy density, radial pressure, and transverse pressure, respectively. Solving the Einstein field Eq.~\eqref{eq. EFE}, one obtains
\begin{align}
	e^{-\lambda}\left(\frac{\nu'}{r}+\frac{1}{r^2}\right)-\frac{1}{r^2} &= 8\pi p, \label{eq. 1st Guv = Tuv}\\
	e^{-\lambda}\left(\frac{1}{2}\nu'' - \frac{1}{4}\lambda'\nu'+\frac{1}{4}(\nu')^2+\frac{\nu'-\lambda'}{2r}\right) &= 8\pi p_t, \label{eq. 2nd Guv = Tuv}\\
	e^{-\lambda}\left(\frac{\lambda'}{r} - \frac{1}{r^2}\right) + \frac{1}{r^2} &= 8\pi \epsilon.
	\label{eq. 3rd Guv = Tuv}
\end{align}
Eq.~\eqref{eq. 3rd Guv = Tuv} can be integrated,
\begin{equation}
	e^{-\lambda(r)} = 1-\frac{2m(r)}{r}, \label{eq. e^lambda}
\end{equation}
with
\begin{equation}
	m(r)\equiv4\pi \int_{0}^{r} \epsilon(\tilde{r}) \tilde{r}^2 d\tilde{r}.
\end{equation}
The function $e^{\nu}$ can then be obtained by substituting Eq.~\eqref{eq. e^lambda} into Eq.~\eqref{eq. 1st Guv = Tuv},
\begin{equation}
	\nu'=\frac{8\pi r^3 p + 2m}{r(r-2m)}. 
	\label{eq. dnu}
\end{equation}
Together with the continuity equation, $\nabla_{\mu}T^{\mu}_{\nu}=0$, the remaining field equations lead us to the Tolman Oppenheimer Volkof (TOV) equation for an anisotropic fluid
\begin{equation}
	\frac{dp}{dr} = -\left(\epsilon+p\right) \frac{m + 4\pi p r^3}{r\left(r - 2m\right)} + 2 ~\frac{p_t - p}{r}. \label{eq. TOV}
\end{equation}

In particular, to solve TOV, we must be supplemented by an {\it equation of state} (EoS); that is, one must assume a relation ansatz between the two out of three quantities: \(\epsilon\), \(p\), and/or \(p_t\). In this study, we choose to define the relation between \(p\) and \(\epsilon\) and then determine the \(p_t\) by solving Eq.~\eqref{eq. TOV}, for which will be discussed in Sec.~\ref{sec. Gravastar construction}.

\subsection{ Hayward-type regular black hole}

Regular black holes, such as the Bardeen~\cite{Bardeen} or the Hayward~\cite{Hayward:2005gi} models, typically assume that the nonsingularity is sourced by a vacuum-like matter that generally satisfies the de Sitter equation of state
\begin{equation}
	p = -\epsilon,
\end{equation}
which satisfies $e^{\nu} = e^{-\lambda}$. The negative pressure causes freely moving particles to behave as if they were repulsed from the center, which avoids further gravitational collapse of matter and prevents the formation of singularity \cite{Gliner:1966}. A medium that supports this kind of equation of state is often called the de Sitter vacuum-like medium, which must be anisotropic if we still consider the TOV Eq.~\eqref{eq. TOV} for perfect fluid, unless in a pure de Sitter `universe' \cite{Cadoni:2022chn}.

Hayward \cite{Hayward:2005gi} proposed a regular spacetime called a minimal model, represented by the metric component
\begin{equation}
	e^{\nu(r)} = e^{-\lambda(r)} = 1 - \frac{2Mr^2}{r^3 + 2l^2M},
\end{equation}
where $M$ is the mass configuration such that $\lim_{r\to\infty}m(r) = M$. The spacetime behaves like Schwarzchild for large $r$ and becomes de Sitter-like around the core:
\begin{equation}
	e^{\nu(r)} \sim
	\begin{cases}
		1 - 2M/r,& r\to\infty\\
		1 - r^2/l^2,& r\to0.
	\end{cases}
\end{equation}
This implies that there is an effective de Sitter length $l$ near $r=0$. The corresponding perfect fluid energy density and pressure that give rise to the corresponding metric are
\begin{eqnarray}
	\epsilon(r) &=& -p(r) = \frac{12l^2M^2}{8\pi\left(r^3 + 2l^2 M\right)^2}, \nonumber\\ p_t(r) &=& \frac{24(r^3 - l^2 M)l^2M^2}{8\pi\left(r^3 + 2l^2 M\right)^3}.
\end{eqnarray}
Since \(p(r) \neq p_t(r)\), the vacuum sourcing Hayward's spacetime geometry is, by definition, anisotropic. Substituting $\ell^3=2l^2M$~\cite{Cadoni:2022chn}, the Hayward metric and energy density can be written as
\begin{equation}
	m(r) = \frac{Mr^3}{r^3+\ell^3},\ \ \ \ \epsilon(r) = \frac{3}{4\pi}\frac{M\ell^3}{\left(r^3 + \ell^3\right)^2}.
	\label{eq. density and mass}
\end{equation}

\begin{figure}[ht!]
	\centering	\includegraphics[width=.5\textwidth]{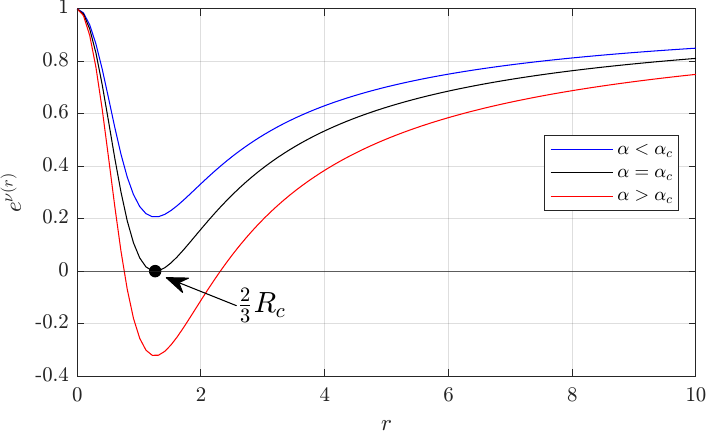}
	\caption{Hayward RBH metrics behavior. For $\alpha>\alpha_c$, the metric yields two distinct roots, corresponding to the presence of two horizons. When $\alpha<\alpha_c$, the solution describes a horizonless configuration. At the critical value $\alpha=\alpha_c$, the solution becomes extremal, featuring a single degenerate horizon.}
	\label{fig:Haywardmetric}
\end{figure}

The metric component becomes
\begin{equation}
	e^{\nu(r)} = e^{-\lambda(r)} = 1 - \frac{2Mr^2}{r^3 + \ell^3}.
	\label{eq. metric comp hayward}
\end{equation}
One can define a quantity
\begin{equation}
	\alpha = \frac{2M}{\ell} = \frac{R_S}{\ell},
\end{equation}
such that $e^{\nu(r)}$ can have one, two, or no horizons depending on the value of $\alpha$. The black hole is extremal at
\begin{equation}
	\alpha_c = \frac{R_c}{\ell} = \frac{3}{\sqrt[3]{4}},
\end{equation} 
which corresponds to the extremal horizon location at $2R_c/3$. The behavior of the metric can be seen in Fig.~\ref{fig:Haywardmetric}.

\subsection{Anisotropic gravastar}
\label{sec. Continuous gravastar}

The original model of gravastar, known as the thin-shell model, consists of three main regions: the interior, the thin-shell, and the exterior, \cite{Mazur:2001fv}
\begin{align}
	&\text{I.}& p &= -\epsilon, && 0<r<r_1\;\;\text{(interior)}\notag\\
	&\text{II.}& p &= \epsilon, && r_1<r<r_2\;\;\text{(thin-shell)}\\
	&\text{III.}& p &= \epsilon=0, && r>r_2\;\;\text{(exterior)}\notag
\end{align}
The negative pressure on the interior region implies that the gravastar has a repulsive de Sitter core, and the thin shell region supports the transition between the de Sitter vacuum interior and the Schwarzschild vacuum on the exterior. 

\begin{figure}[ht!]
	\centering
	\includegraphics[width=.5\textwidth]{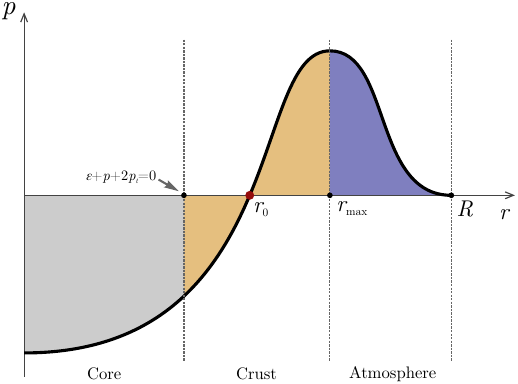}
	\caption{Qualitative sketch of anisotropic gravastar's  pressure profile proposed by \cite{Cattoen:2005he}.}
	\label{fig.anisotropic gravastar pressure}
\end{figure}

However, the discontinuous jump from negative to positive pressure is not physically viable. Cattoen et al. \cite{Cattoen:2005he} give an idea of gravastar with continuous pressure throughout the entire region of the star as shown in Fig.~\ref{fig.anisotropic gravastar pressure}. The idea is to introduce an anisotropic medium as the source of matter. The pressure profile must satisfy the following requirements:
\begin{enumerate}[nolistsep]
	\item the de Sitter core requires $p(r=0) = p_t(r=0) = -\epsilon(r=0)$,
	\item there must be two radial points where the $p$ vanishes: at $r=r_0$ and $r=R$; $r_0$ is the transition point between the negative pressure of de Sitter vacuum and positive pressure of the atmosphere, and $R$ is the surface of the star,
	\item $p'(r=0) =0$ to maintain the regularity.
\end{enumerate}
Some restrictions were also given regarding the energy conditions. The null energy condition (NEC) given by $\epsilon + p \geq 0$ and $\epsilon + p_t \geq 0$ together with the weak energy condition (WEC) $\epsilon \geq 0$ are always satisfied throughout the entire region of the star. The strong energy condition (SEC) given by $\epsilon + p + 2p_t \geq 0$ is indeed violated near the core of the star, and the dominant energy condition (DEC) $|p|\leq\epsilon$ and $|p_t|\leq \epsilon$ is allowed to be violated.

The gravastar interior itself has three main regions:
\begin{enumerate}[nolistsep]
	\item \textit{the core}, located from the star center $r=0$ to a point where the SEC satisfied,
	\item \textit{the crust}, located from the outer side of the core to the point of maximum positive radial pressure $r=r_{max}$ where $dp/dr=0$,
	\item \textit{the atmosphere}, located outside the crust to the star surface $R$.
\end{enumerate}

DeBenedictis, {\it et al.}~\cite{DeBenedictis:2005vp} realized Cattoen's idea by introducing an ansatz of anisotropy which is related to the local compactness;
\begin{equation}
	\Delta(r) = \frac{p_t(r) - p(r)}{\epsilon(r)} = \sigma \frac{m(r)}{r}
	\label{eq. anisotropy ansatz},
\end{equation}
where $\Delta(r)$ is the measure of anisotropy and $\sigma$ is a constant that can be constrained by the energy conditions. Choosing an appropriate energy density profile $\epsilon$, one can numerically solve the TOV equation for anisotropic fluids, which now reads
\begin{equation}
	\frac{dp}{dr} = -\left(\epsilon+p\right) \frac{m + 4\pi p r^3}{r\left(r - 2m\right)} + 2 \epsilon\frac{\Delta}{r}, \label{eq. TOV delta}
\end{equation}
to obtain the radial pressure $p$ by imposing the initial condition $p(r=0)=-\epsilon(r=0)$. The transverse pressure $p_t$ can then be calculated using the ansatz of anisotropy. Alternatively, one can impose an equation of state \(p=p(\epsilon)\) and determine the transverse pressure (and simultaneously, the anisotropy measure) by solving the TOV equation of Eq.~\eqref{eq. TOV}.

%=======================================================================
\section{Transition between regular black hole and anisotropic gravastar}
\label{sec. Transition Construct}

In this section, we try to construct a star model as the horizonless configuration of a regular black hole's spacetime that satisfies the anisotropic gravastar requirement. We choose Hayward's minimal model with the form of energy density and mass profile described by Eq.~\eqref{eq. density and mass}.

\subsection{Construction of anisotropic gravastar model}
\label{sec. Gravastar construction}

For convenience, we define a dimensionless variable
\begin{equation}
	x\equiv\frac{\alpha}{\alpha_c}.
\end{equation}
For $x<1$, the object corresponds to a star configuration, while for $x\geq1$, it becomes a regular black hole. The event horizon characterizes the radius of a black hole, while the radius of a star is defined by the point where the pressure drops to zero. Therefore, we propose an explicit definition for the star radius $R$,
\begin{equation}
	R \equiv \frac{\alpha_c}{\alpha} R_c = \frac{R_c}{x}.
	\label{eq:star_radius}
\end{equation}
They indicate that the star radius is related to the critical Schwarzschild radius $R_c$. With this definition, one can easily see that the dimensionless variable \(x\) corresponds to the `mass' configuration for a fixed value of \(\ell\), where \(M=x\alpha_c \ell/2\). Thus, Eq.~\eqref{eq:star_radius} shows that the object will become more compact for more extensive mass configuration. We force the energy density to vanish at the surface so that \(\epsilon(r\geq R)=0\). The density and mass profile of Eq.~\eqref{eq. density and mass} then can be written as
\begin{equation}
	\epsilon(r) = \frac{3}{8\pi} \frac{x\alpha_c\ell^4}{\left(r^3 + \ell_p^3\right)^2},\qquad
	m(r) = \frac{x\alpha_c\ell r^3}{2\left(r^3 + \ell^3\right)}.
	\label{eq. density and mass in x}
\end{equation}

The anisotropic ansatz of Eq.~\eqref{eq. anisotropy ansatz} somehow implies that the radial pressure is proportional to the energy density multiplied by the local compactness, i.e.,
\begin{equation}
	p(r) \propto \epsilon(r) \frac{m(r)}{r}.
\end{equation}
Inspired by this relation, we propose a form of pressure profile that is related to local compactness \(m(r)/r\);
\begin{equation}
	p(r) = -\epsilon(r)\left[1-\beta(r)\frac{m(r)}{r}\right],
\end{equation}
where \(\beta\) is a quantity with dimension of $\text{\it radius}\times\text{\it mass}^{-1}$. Since we have already defined the radius explicitly and we require (continuous) zero pressure at the star radius, we conclude that $\beta(r)$ must be proportional to the inverse of total compactness, i.e.
\begin{equation*}
	\beta(r) \propto \frac{R}{m(R)},
\end{equation*}
for which can be multiplied by another dimensionless function $g(r)$ satisfying
\begin{equation}
	g(r) =\begin{cases}
		\geq 0,& 0\leq r<R,\\
		1,& r = R.
	\end{cases}
	\label{eq. req g(r)}
\end{equation}
Hence, the pressure profile can be written explicitly as
\begin{equation}
	p(r) = -\epsilon(r)\left[1-g(r)\frac{R}{r}\frac{m(r)}{m(R)}\right].
\end{equation}
With this particular pressure profile, it can be seen that at $r=R$, the last term becomes unity, leading to the pressure becoming zero, thus defining the star radius. At the center ($r=0$), the local compactness is zero, resulting in $p=-\epsilon$, indicating that we have a de Sitter core as an anisotropic gravastar requires. It can also be observed that the condition $p'(r=0)=0$ is always satisfied, regardless of the specific form of $g(r)$. Depending on the choice of $g(r)$, we can achieve other desired features of an anisotropic gravastar as discussed previously in Sec.~\ref{sec. Continuous gravastar}, which will be elaborated in Sec.~\ref{sec. model results}.

\begin{figure}[ht!]
	\centering
	\includegraphics[width=0.42\textwidth]{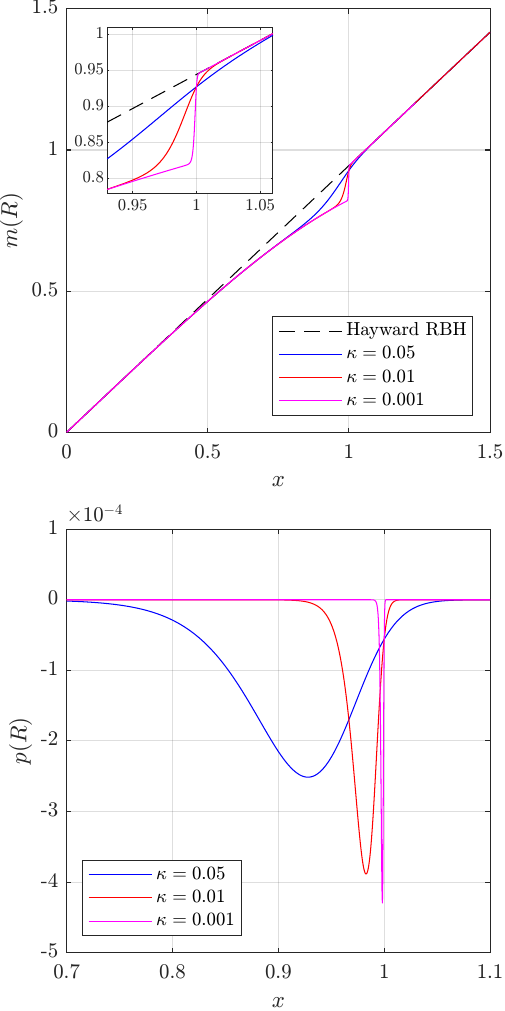}
	\caption{The total mass $m(R)$ and surface pressure $p(R)$ as a function of $x$ for several value of $\kappa$.}
	\label{fig. mass pressure x gconst}
\end{figure}

However, the corresponding pressure profile does not transition smoothly to a regular black hole as $x\to1$. Regular black hole solutions require $p=-\epsilon$ and do not have a well-defined radius. This indefinite radius must be considered to ensure that the total mass $m(R)$ has a continuous value for every value of $x$. Thus, we introduce a cut-off function $\Theta(x-1)$ which satisfies
\begin{equation}
	\Theta(x-1)=
	\begin{cases}
		1,& x < 1\\
		0,& \text{otherwise},
	\end{cases}
\end{equation}
to modify the radius and the radial pressure profile as follows:
\begin{eqnarray}
	R &=& \frac{R_c}{x} \left[\Theta(x-1)\right]^{-1},\nonumber\\ p(r) &=& -\epsilon(r)\left[1-g(r)\frac{R}{r}\frac{m(r)}{m(R)}\Theta(x-1)^n\right],
\end{eqnarray}
where $n>1$ is an integer chosen to ensure that the cut-off function in the pressure profile is more dominant. As $x\to1$, the radius becomes indefinite, and the cut-off function in the pressure profile does not vanish if $n>1$. This ensures that the term inside the brackets in the pressure profile becomes zero as $x\to1$, resulting in the de Sitter EoS $p=-\epsilon$. For simplicity, we will use $n=2$. We choose the cut-off function to be the Sigmoidal jump function
\begin{equation}
	\Theta(x-1) = \frac{1}{1+e^{(x-1)/\kappa}},
\end{equation}
where $\kappa$ determines the \textit{smoothness} of the transition. This smoothness defines a \textit{transition zone} between the anisotropic gravastar and the regular black hole. To investigate this, we plot the relation between $x$ and the resulting total mass $m(R)$ on the left side of Fig.~\ref{fig. mass pressure x gconst}. Based on the results, we choose $\kappa=0.01$ to achieve sufficient transition smoothness.

In general, this model consists of two regions: the interior and the exterior. The geometry of the exterior region is described by Schwarzschild metric since we have $\epsilon(r\geq R)=p(r\geq R)=0$, so that
\begin{equation}
	e^{\nu(r)}_{ext} = e^{-\lambda(r)}_{ext} = 1 - \frac{2m(R)}{r}= 1 - \frac{x\alpha_c\ell R^3}{r\left(R^3 + \ell^3\right)},\qquad r\geq R.
\end{equation}
On the other hand, the interior region is described by Eqs.~\eqref{eq. e^lambda} and~\eqref{eq. dnu},
\begin{eqnarray}
	e^{-\lambda(r)}_{int} &=& 1 - \frac{x\alpha_c\ell r^2}{r^3 + \ell^3},\nonumber\\ e^{\nu(r)}_{int} &=& \exp\left[2\int_{R}^{r} \frac{4\pi \tilde{r}^3 p(\tilde{r}) + m(\tilde{r})}{\tilde{r}(\tilde{r}-2m(\tilde{r}))}d\tilde{r}\right],\qquad r<R.\nonumber\\
	\label{eq. metric gravastar}
\end{eqnarray}
The integration of Eq.~\eqref{eq. metric comp hayward} is carried numerically with backward integration from $r=R$ by setting the initial condition on the surface to match the interior solution with the Schwarzschild exterior, i.e., $e^{\nu(R)}_{int}=e^{\nu(R)}_{ext}$. However, if $x\geq1$, then the spacetime becomes Hayward-type everywhere described by Eq.~\eqref{eq. metric comp hayward} since we will have indefinite $R$ and the de Sitter equation of state $p=-\epsilon$.

Since we aim to study the shadow image of this particular model, we can constrain the range of values for $x$ such that it produces a photon sphere outside the star radius. With a Schwarzschild exterior, the photon sphere radius is at $R_{ps} = 3M$. We find that the critical value is
\begin{equation}
	x_{ps} \approx 0.85324.
\end{equation}
Thus, the star surface will resides below the photon sphere if $x > x_{ps}$.

\subsection{Model results}
\label{sec. model results}

Here, we provide examples of two models of the pressure profile with different forms of $g(r)$. The first model corresponds to a constant $g(r)$, while in the second model, we consider a simple function that satisfies the requirement given in Eq.~\eqref{eq. req g(r)}.

\subsubsection{Model 1}
\begin{figure*}[htbp!]
	\centering
	\includegraphics[width=0.9\textwidth]{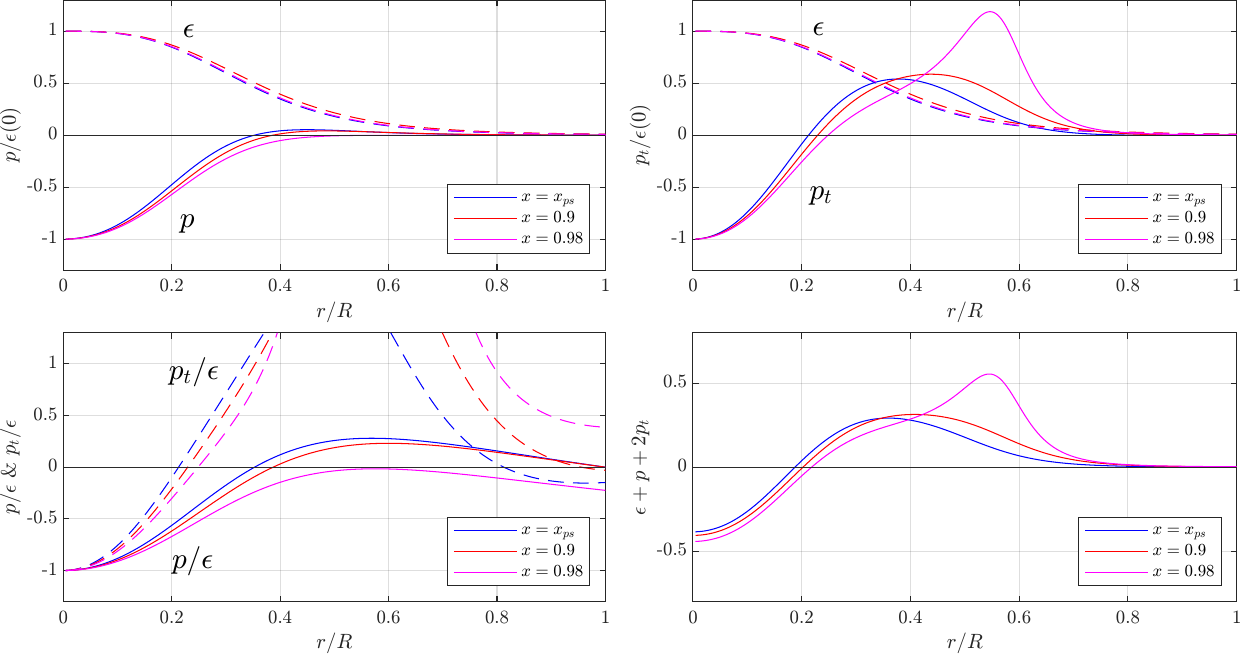}
	\caption{Radial (top left) and transverse (bottom right) pressure profile, and energy conditions (bottom left and right) for \(g(r)=1\). In the top two graph, the dashed line represent the density profile \(\epsilon(r)/\epsilon(0)\). We choose \(\kappa = 0.01\).}
	\label{fig. rpressure qpressure gconst}
\end{figure*}

\begin{figure}[htbp!]
	\centering
	\includegraphics[width=0.42\textwidth]{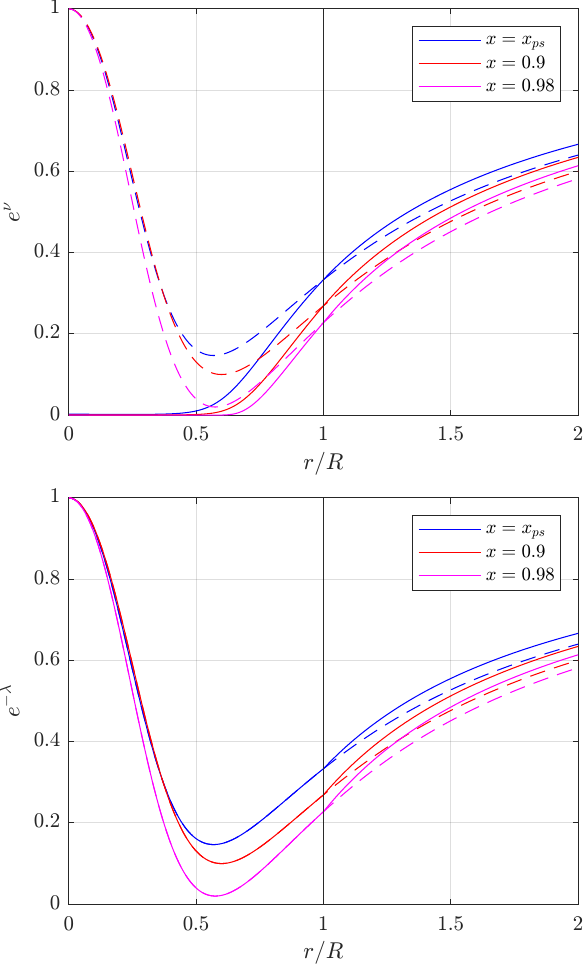}
	\caption{The \(e^{\nu}\) (top) and \(e^{-\lambda}\) (bottom) metric component for \(g(r)=1\), presented by colored solid lines. The dashed lines shows the Hayward metric component from Eq.~\eqref{eq. metric comp hayward}, and the black solid line indicates the star radius location.}
	\label{fig. metric gconst}
\end{figure}

For the first model, we choose a simple form of the dimensionless function, specifically $g(r) = 1$. This makes it a constant and still fulfills the requirement of Eq.~\eqref{eq. req g(r)}. The pressure profile is explicitly written as
\begin{equation}
	p(r) = -\epsilon(r) \left[1 - \frac{R}{r}\frac{m(r)}{m(R)}\Theta(x-1)^2\right].
\end{equation}
The resulting pressure profile and its energy conditions for this model are presented in Fig.~\ref{fig. rpressure qpressure gconst}, while the metric solutions are shown in Fig.~\ref{fig. metric gconst}.

The requiement for anisotropic gravastar is clearly achieved within a certain value of $x$, where we obtain a negative and positive radial pressure region. From the bottom left figure of Fig.~\ref{fig. rpressure qpressure gconst}, it can be seen that the transverse pressure exceeds the energy density at some region, indicating the violation of dominant energy condition. It can also be observed from the bottom right figure that the strong energy condition is only violated at the core of the star.

The EoS\textemdash as stated in Eq.~\eqref{eq. metric gravastar}\textemdash affects the $e^{\nu}$ metric component, while the $e^{\lambda}$ component only depends on the mass function. By matching the solution with the Schwarzshcild exterior component, the integration of Eq.~\eqref{eq. metric gravastar} results in a near-zero value of the $e^{\nu}$ component inside the interior, notably near the core region. At first glance, this was counter-intuitive since we expect that the metric value would approach unity at the center because, at that point, we have \(p=-\epsilon\). However, this is also the case for the thin-shell gravastar solution, where the $e^{\nu}$ component does not approach unity if the positive pressure region, i.e., the thin-shell, gives a contribution to the total mass of the gravastar \cite{Rosa:2024bqv}. Hence, it is reasonable for our gravastar model to produce a near-zero value of the $e^{\nu}$ component near the core region since we have a thicker positive pressure region and a quite significant contribution to the mass of the star\footnote{It should be noted that this argument does not provide the whole picture, since we have a more extreme near-zero $e^{\nu}$ value with $x=0.98$. At the same time, it does not contain a positive matter region.}. The $e^{\nu}$ component is similar to the quantum horizonless compact object described in \cite{Chen:2024ibc}.

\begin{figure}[htbp!]
	\centering
	\includegraphics[width=.7\textwidth]{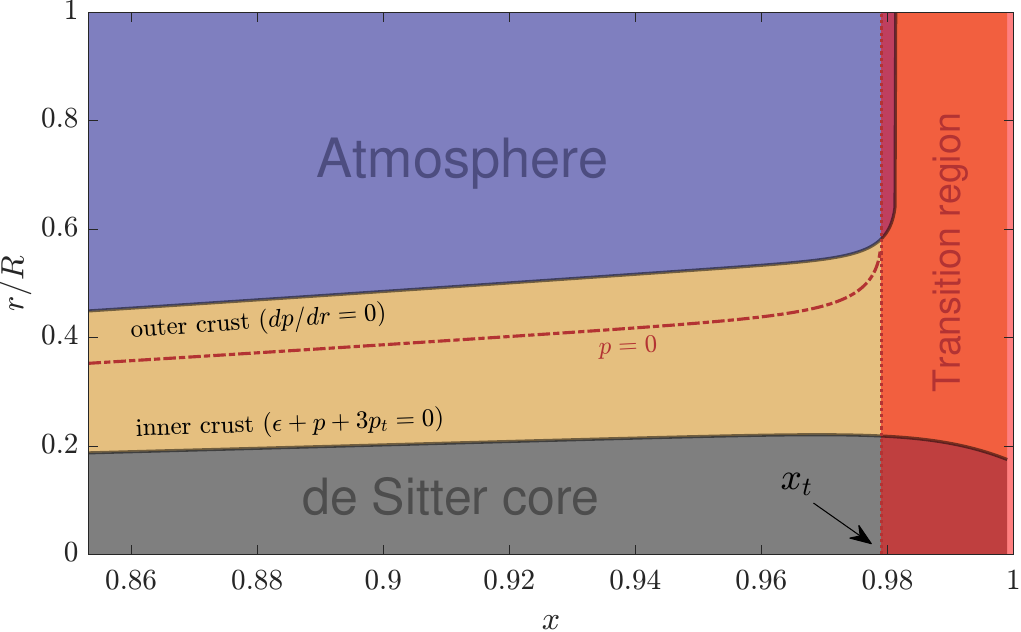}
	\caption{Inner structure of the corresponding proposed gravastar model with $g(r) = 1$ and $\kappa = 0.01$, where $x_t$ denotes the transition zone boundary.}
	\label{fig. grav region gconst}
\end{figure}

To investigate the region of interest (\textit{core, crust,} and \textit{atmosphere}) inside the interior, we plot the boundary of each region for every value of $x$ ranging from $x=x_{ps}$ to $x=1$, as shown in Fig.~\ref{fig. grav region gconst}. The dot-dashed lines indicate the transition point of positive and negative pressure region, $r_0$, where $p(r_0) = 0$. The $r_0$ is undefined at some point of $x$, which leads to a negative pressure in the entire region of the star. We define the region of $x$ with undefined $r_0$ as the \textit{transition region}. We found that the \textit{size} of this region depends on the value of $\kappa$ and the function $g(r)$, as seen in Fig.~\ref{fig. transition region x}. The appearance of this transition region is one consequence of our approach, especially the use of the cut-off $\Theta$ function.

\subsubsection{Model 2}
We consider a simple form of $g(r)$;
\begin{equation}
	g(r)=\left[1+\omega\left(1-\frac{r}{R}\right)\right]^{-1},
	\label{eq. g(r) model 2}
\end{equation}
with $\omega$ is a constant that can be adjusted or constrained. We see that this function satisfies the requirement of Eq.~\eqref{eq. req g(r)}. The pressure profile can then be expressed as
\begin{equation}
	p(r) = -\epsilon(r) \left[1 - \left(1+\omega\left(1-\frac{r}{R}\right)\right)^{-1}\frac{R}{r}\frac{m(r)}{m(R)}\Theta(x-1)^2\right].
\end{equation}
The free constant $\omega$ generalize the constant $g(r)$ model given in the previous section, where one can obtain $g(r)=1$ if $\omega=0$. We also see that it is possible to obtain the de Sitter equation of state $p=-\epsilon$ as $\omega \to \infty$. However, we found that a particular value of $\omega$ that is higher than a certain value (which will be discussed shortly) would produce an entirely negative pressure inside the region of the star with $x$ value less than unity.

\begin{figure*}[ht!]
	\centering
	\includegraphics[width=0.9\textwidth]{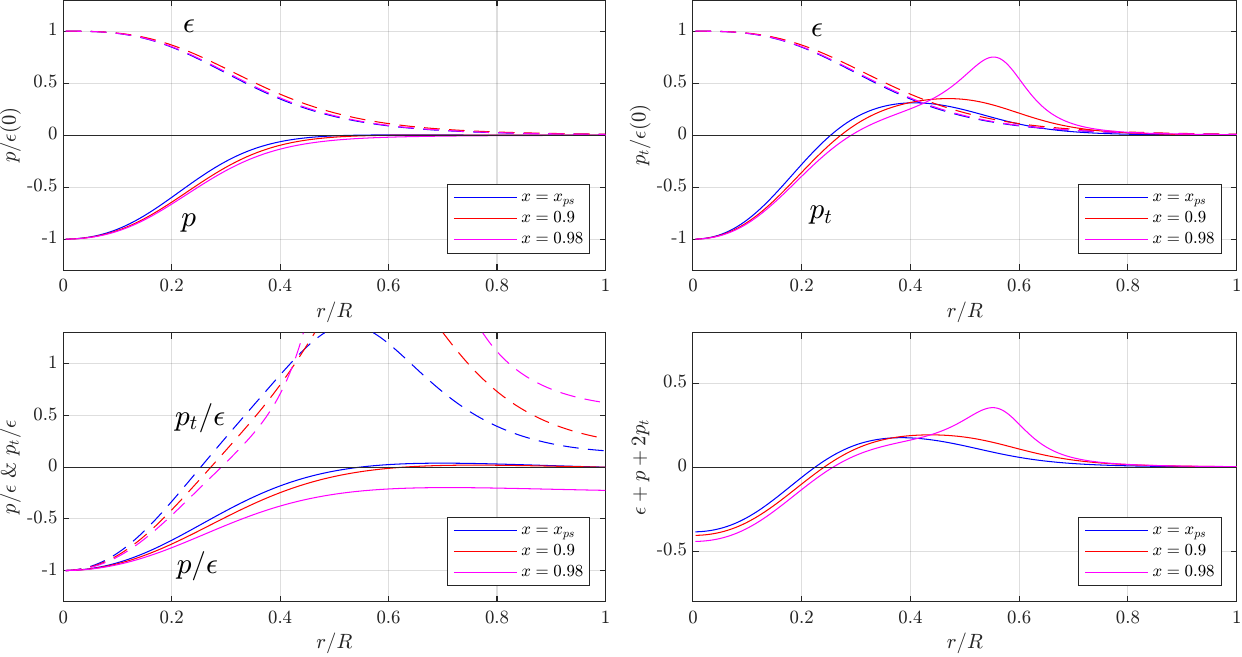}
	\caption{Same description as Fig.~\ref{fig. rpressure qpressure gconst} with \(g(r)=\left[1+\omega\left(1-r/R\right)\right]^{-1}\).}
	\label{fig. rpressure qpressure gfunc}
\end{figure*}

In this study, we determine the value of $\omega$ such that the radial pressure becomes negative everywhere as $x\to1$ while disregarding the jump function $\Theta(x-1)$. It is characterized by $p(R)= p'(R) = 0$, and such condition will be satisfied if\footnote{
	This condition comes from the fact that there should be only one pressure zero, at \(r=R\), so that the maximum radial pressure value is zero and located at the surface.
}
\begin{equation}
	\left.\frac{g'(r)}{g(r)}\right|_{r=R} = \frac{\alpha_c^3-2}{\alpha_c\ell\left(\alpha_c^3 + 1\right)}.
\end{equation}
With our form of $g(r)$ given by Eq.~\eqref{eq. g(r) model 2}, we find the value of $\omega$ to be
\begin{equation}
	\omega = \frac{\alpha_c^3-2}{\alpha_c^3 + 1}.
\end{equation}
This approach allows for a smooth transition to the de Sitter interior's negative pressure, minimizing the jump function's influence. The results are presented in Fig.~\ref{fig. rpressure qpressure gfunc}. 

\begin{figure}[htb!]
	\centering
	\includegraphics[width=0.42\textwidth]{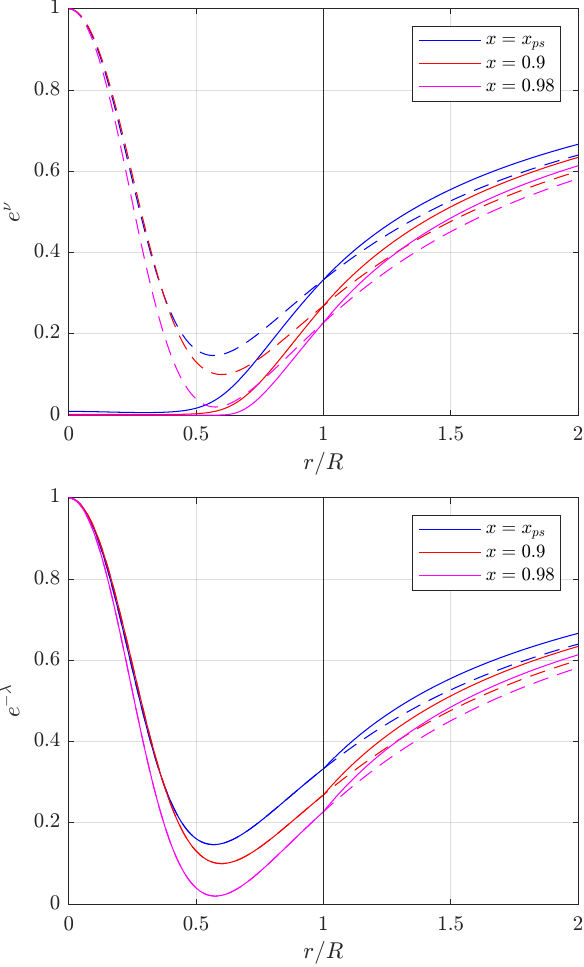}
	\caption{Same description as Fig.~\ref{fig. metric gconst} with \(g(r)=\left[1+\omega\left(1-r/R\right)\right]^{-1}\).}
	\label{fig. metric gfunc}
\end{figure}

This model, with the given $\omega$ value, produces a lower pressure at the outer crust boundary and it shifts to the outer side region. The violation of the dominant energy condition is also relaxed, where the peak of the transverse pressure is significantly reduced. The $r_0$ point is also shifted to the outer side, producing a smaller positive pressure region as shown in Fig.~\ref{fig. metric gfunc}. The $e^{\nu}$ component value near the core region is slightly higher than what we obtain on the previous model, partially confirming our argument on the contribution of mass on the positive pressure region.

\begin{figure}[htb!]
	\centering
	\includegraphics[width=.7\textwidth]{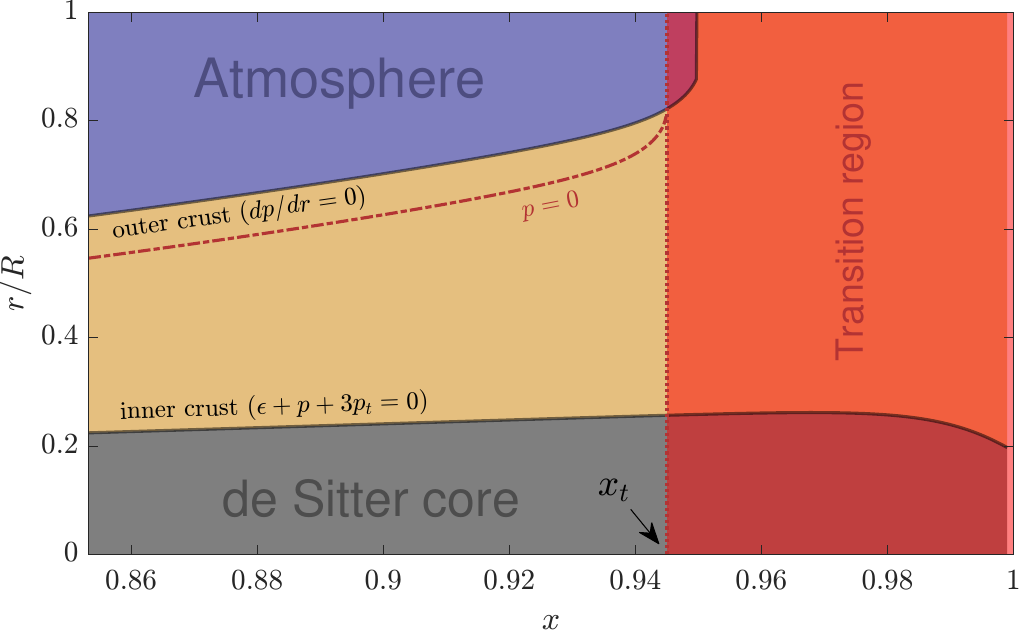}
	\caption{Similar plot as Fig.~\ref{fig. grav region gconst} with $g(r)=\left[1+\omega\left(1-r/R\right)\right]^{-1}$.}
	\label{fig. grav region gfunc}
\end{figure}

We also plot the boundary of each region for this particular model in Fig.~\ref{fig. grav region gfunc}. We can conclude that the $g(r)$ plays a significant role in the region boundaries so that one can find a more desirable function to satisfy a desired condition. It also can be seen that the transition region is wider, which comes from the fact that this particular $g(r)$ gives an entirely negative pressure throughout the interior region of the star as $x\to1$, so that the effects of $\Theta(x-1)$ kicks in early.

\begin{figure}[htb!]
	\centering
	\includegraphics[width=.7\textwidth]{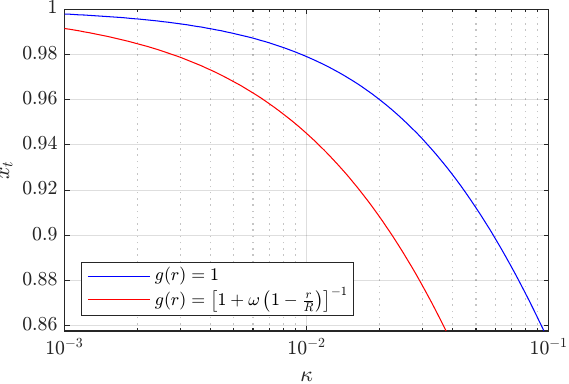}
	\caption{Impact of $g(r)$ profile on transition zone boundary $x_t$ as a function of $\kappa$.}
	\label{fig. transition region x}
\end{figure}

As previously mentioned, the transition region boundary also depends on the smoothness of the jump function, i.e., the value of $\kappa$. From Figs.~\ref{fig. grav region gconst} and \ref{fig. grav region gfunc}, the boundary are also differ for a different $g(r)$ function. Thus, we plot the transition boundary as a function of $\kappa$ for each $g(r)$ function in Fig.~\ref{fig. transition region x}. As expected, the transition region boundary gets closer to $x=1$ as we reduce the jump function's smoothness, i.e., smaller $\kappa$.

\subsubsection{Strong redshifts inside the star}
\begin{figure}[ht!]
	\centering
	\includegraphics[width=0.7\textwidth]{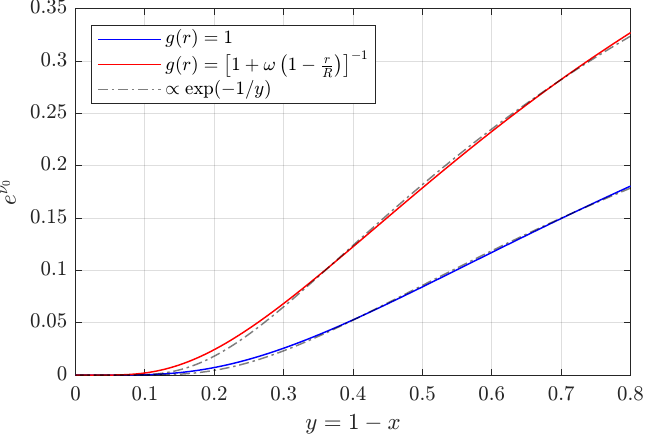}
	\caption{The $e^{\nu}$ metric component at $r=0$, representing the strong redshift effects near the central core. Curve fitting to a function $\propto \exp(-1/y)$ are provided with the gray dash-dotted line for both form of $g(r)$.}
	\label{fig. central redshift}
\end{figure}

As shown in Figs.~\ref{fig. metric gconst} and \ref{fig. metric gfunc}, the interior $e^{\nu}$ component gives rise to strong gravitational redshifts, similar to those observed in the QHCO model of Ref.~\cite{Chen:2024ibc}. It can be seen, particularly in Fig.~\ref{fig. metric gfunc}, that higher redshifts are produced at larger values of $x$. Therefore, here we will investigate the strong redshift effect inside the star's core, specifically around $r=0$. We define a new quantity
\begin{equation}
    y = 1-x.
\end{equation}
which can be thought as a quantity related to strong redshifts. Since we're interested in the horizonless solution with $0>x>1$, the value of $y$ will be ranging only from zero to unity. However, zero value of $x$ implies either infinite $\ell$ or zero mass, so we only consider $0.2<x<1$ which corresponds to $0<y<0.8$. We plot the central time metric component $e^{\nu_0} \equiv e^{\nu(0)}$ with respect to $y$, and the results are shown in Fig.~\ref{fig. central redshift}.

We can clearly see that the decrease in $y$ parameter significantly surpresses the central $e^\nu$ component with non-linear behavior approximately proportional to $\sim\exp(-1/y)$. As a consequence, photons emitted near the star's core will experience strong redshift and a substantial (but finite) time delay. A similar behavior occurs in the QHCO discussed in Ref.~\cite{Chen:2024ibc}, where smaller Hawking-like radiation and interaction parameters (say, $\eta$) lead to increased strong redshifts with the factor of $\propto\exp(-1/\eta)$. Hence, we can roughly conclude that there is a correlation between $y$ in our model with the parameters introduced in the QHCO, particularly with the interaction parameter.

Strong redshift is widely known to significantly contribute to the image darkening mechanism for light sources located within the corresponding region through Lorentz invariance \cite{Chen:2024ibc,Rosa:2024bqv,Rosa:2023qcv,Rosa:2023hfm}, which will be discussed shortly in Sec.~\ref{sec. Thin accretion disk}. In Ref.~\cite{Chen:2024ibc}, the authors also demonstrated a mechanism by which time delay contributes to further image darkening. However, our study will not address this latter aspect.

%=======================================================================
\section{Geodesic and shadow image}
\label{sec. Geodesic and shadow image}

This study will compare the optical observation results between the Hayward-type regular black hole and our proposed anisotropic gravastar model by analyzing their shadow images. To obtain a simplified image model, many studies of compact objects and black hole shadows use an optically and geometrically thin accretion disk as the observation signature. The image models are generated through a ray tracing procedure involving null geodesic calculations, which will be discussed in this section. For convenience, let us re-express the metric Eq.~\eqref{eq. SSS ansatz enu} by
\begin{equation}
	ds^2 = -A(r)dt^2 + B(r)dr^2 + r^2d\Omega^2.
	\label{eq. SSS ansatz AB}
\end{equation}

\subsection{Geodesic motion and ray tracing}

The test particle trajectory is described by the geodesic equations
\begin{equation}
	\frac{dk^{\mu}}{d\tau} + \Gamma^{\mu}_{\alpha\beta} \frac{dx^{\alpha}}{d\tau} \frac{dx^{\beta}}{d\tau} = 0,
\end{equation}
where $\tau$ is an affine parameter, $x^{\mu} = (t,r,\theta,\phi)$, and $k^{\mu} = dx^{\mu}/d\tau = \dot{x}^{\mu}$. These equations are constrained by the 4-velocity condition,
\begin{eqnarray}
	-A(r)\dot{t}^2 &+& B(r) \dot{r}^2 + r^2 \dot{\theta}^2 \nonumber\\ &+& r^2\sin^2{\theta}\dot{\phi}^2 =
	\begin{cases}
		-1, & \text{timelike/massive}\\
		0. & \text{null}
	\end{cases}
	\label{eq. four velocity condition}\nonumber\\
\end{eqnarray}
With the static and spherically symmetric ansatz \eqref{eq. SSS ansatz AB}, and by restricting the geodesic motion to the equatorial plane $\theta = \pi/2, \dot{\theta} = 0$, we obtain two conserved quantities:
\begin{equation}
	A(r)\dot{t} = E, \qquad \dot{\phi} r^2 = L. \label{eq. conserved quantity geodesic}
\end{equation}

\begin{figure}[ht!]
	\centering
	\includegraphics[width=0.7\textwidth]{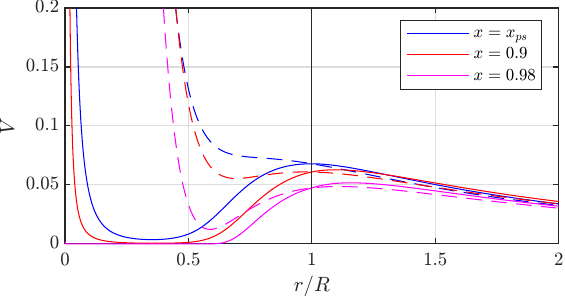}
	\caption{Photon effective potential for our gravastar model (solid lines) and Hayward spacetime (dashed lines) with several value of $x$.}
	\label{fig. photon eff potential}
\end{figure}

For photon trajectories, substituting Eq.~\eqref{eq. conserved quantity geodesic} into Eq.~\eqref{eq. four velocity condition} for the null condition yields
\begin{equation}
	\label{photonnullcondition}
	\frac{d\phi}{dr} = \frac{1}{r^2}\sqrt{\frac{A(r)B(r)}{1/b^2 - V(r)}},
\end{equation}
where $V(r)$ is the photon effective potential, defined as
\begin{equation}
	V(r) \equiv \frac{A(r)}{r^2}.
\end{equation}
The constant $b\equiv L/E$ is known as the impact parameter. The value of $L$ and $E$ can be determined from initial conditions
\begin{equation}
	E = \sqrt{A(r_0)\left[B(r_0) \dot{r}_0^2 + r^2_0 \dot{\phi}_0^2\right]},\qquad L = \dot{\phi}_0r_0^2.
\end{equation}
We plot the photon effective potential in Fig.~\ref{fig. photon eff potential}.

To solve~\eqref{photonnullcondition}, we perform integration in two parts: inward and outward. First, we integrate the null geodesic inward by setting the step size $\Delta r$ to a negative value, starting from an observer at $r_0$. Once the integration reaches the turning point where $V(r)\geq1/b^2$, the integration direction reverses, and $\Delta r$ is set to a positive value. The other initial condition value can be determined by the initial direction of the light ray from the observer, given by the ratio $\dot{\phi}_0/\dot{r}_0$. It is natural to use the initial $\Delta r$ for $\dot{r}_0$ value\footnote{
	The value of $\Delta r$ does not need to be constant if we use an adaptive step-size method, allowing the initial $\Delta r$ to differ from subsequent $\Delta r$.
}.

\subsection{Thin accretion disk}
\label{sec. Thin accretion disk}

We assume an optically and geometrically thin accretion disk with monochromatic emission. This assumption is common in several studies of black hole shadows (e.g. \cite{Gralla:2020srx,Gralla:2019xty,Carballo-Rubio:2022aed,Zeng:2023fqy,Meng:2023htc,Gao:2023mjb}) and other ultracompact objects (e.g. \cite{Chen:2024ibc,Rosa:2024bqv,Rosa:2023qcv,Guerrero:2022msp}). An optically thin accretion disk is transparent, allowing light to pass through without significant blocking or attenuation. The geometrically thin property assumes the disk is infinitely thin, spanning only a two-dimensional plane. To model a more realistic accretion disk, we will also consider the effects of accretion flow and gravitational redshift, which will be discussed shortly.

\subsubsection{Emission profile}

It is usually expected that the accumulation of matter on the accretion disk leads to electromagnetic emission of the accretion disk. For simplicity, we consider an accretion disk emission profile $I_e(r)$, which is related to the matter density and the temperature of the accretion disk. We adopt the emission profile known as the Gralla-Lupsasca-Marrone (GLM) model \cite{Gralla:2020srx},
\begin{equation}
	I_e(r) = \frac{\exp\left\{-\frac{1}{2}\left[\gamma + \operatorname{arcsinh}{\left(\frac{r-\mu}{\sigma}\right)}\right]^2\right\}}{\sqrt{\left(r-\mu\right)^2+\sigma^2}},
	\label{eq. intensity profile glm}
\end{equation}
where $\gamma$, $\mu$, and $\sigma$ are some parameters that determines the shape of the accretion disk. In this study, we consider two types of emission profiles:
\begin{enumerate}[nolistsep]
	\item \textbf{GLM1}; characterized by $\gamma = -2$, $\mu = R_{ISCO}$, and $\sigma = M/4$, with $R_{ISCO}$ is the radius of the innermost stable circular orbit (ISCO) for a massive particle. This type of accretion disk assumes that the emission peaks and stops near the ISCO radius. This model is considered because any massive particle will eventually fall towards the center if it passes within the ISCO radius, leading to peak matter density around the ISCO radius.
	
	\item \textbf{GLM2}; characterized by $\gamma=\mu=0$ and $\sigma = 2M$. For this emission profile, we assume that the accretion disk spans through the center of the object, as there is no restriction for matter in horizonless spacetime to reach the center. The emission peaks at the center, $r=0$. However, this type of accretion disk will not be considered for inclined observations.
\end{enumerate}

\begin{figure}[ht!]
	\centering
	\includegraphics[width=0.7\textwidth]{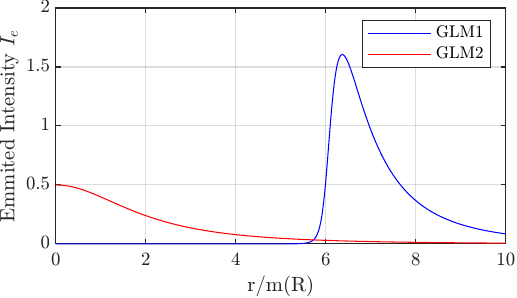}
	\caption{Emission profile for GLM1 (blue) and GLM2 (red) in arbitrary units. In this graph, we use the example of Schwarzschild ISCO radius $R_{ISCO} = 6m(R)$.}
	\label{fig. intensity profile}
\end{figure}

Both emission profiles are illustrated in Fig.~\ref{fig. intensity profile}. The observed emission profile will be influenced by the relative accretion flow and redshift factor, which will be discussed in the following sections.

\subsubsection{Innermost stable circular orbit (ISCO) and redshift factor}
Since we have Schwarzschild geometry outside the gravastar configuration, we can choose \(R_{ISCO}=6m(R)\). However, for the general form of spacetime, the \(R_{ISCO}\) can be calculated numerically by solving~\cite{Gao:2023mjb}
\begin{equation}
	\left.3A(r)A'(r) - 2rA'(r)^2 + rA(r)A''(r)\right|_{r=R_{ISCO}} = 0
	\label{eq. R Isco determine}
\end{equation}

We incorporate the motion of the accretion flow to obtain a more realistic shadow image. The accretion flow affects the observed intensity of the disk due to the relativistic Doppler effect of incoming and outgoing matter relative to the observer. We assume Keplerian accretion flow motion on the equatorial plane, with its four-velocity expressed as
\begin{equation}
	u^{\mu}_e = (u^t,u^r,0,u^{\phi}).
	\label{eq. four velocity accretion}
\end{equation}
Outside the ISCO radius, we assume the accretion flow does not experience any radial drift, so $u^r=0$. With the spacetime metric in form of Eq.~\eqref{eq. SSS ansatz AB}, the $u^{\phi}$-component is given by~\cite{Ryan:1995wh,Ozel:2021ayr}
\begin{equation}
	u^{\phi}_e(r) = \frac{\sqrt{A'(r)}}{\sqrt{2A(r)r - A'(r)r^2}}.
\end{equation}
The $u^{t}$-component can be obtained by the four-velocity condition of Eq.~\eqref{eq. four velocity condition} for timelike particles:
\begin{equation}
	u^{t}_e(r) = \frac{\sqrt{2}}{\sqrt{2A(r) - A'(r)r}}.
\end{equation}
The observed photon frequency $\nu_o$ is weighted to the emitted frequency $\nu_e$ by the energy correction factor $\tilde{g}$ written as \cite{Ozel:2021ayr,M:2022pex,Kumaran:2023brp}
\begin{equation}
	\nu_o = \tilde{g} \nu_e, \qquad
	\tilde{g} = \frac{-k_{\nu}u^{\nu}_o}{-k_{\mu}u^{\mu}_e},
	\label{eq. energy corr f}
\end{equation}
where $k_{\mu} = g_{\mu\nu}k^{\nu}$ is the four-velocity of the light ray and $u^{\nu}_o$ is the four-velocity of the observer. Since we assume that the far away observer is at rest in a flat spacetime, we have $u^{\nu}_o = (1,0,0,0)$ so that
\begin{equation}
	\tilde{g} = \left(u^t_e - \frac{k_{\phi}}{E}u^\phi_e\right)^{-1}.
\end{equation}
If we consider a static accretion disk where $u^{\phi}_e = 0$, we will have $\tilde{g} = 1/u^t_e = \sqrt{A(r)}$. In this case, it only calculate the effects of gravitational redshift which has been used by several authors to calculate such effects on the axial observation images, e.g. in \cite{Rosa:2024bqv,Rosa:2023qcv,Rosa:2023hfm,Zeng:2023fqy,Meng:2023htc}. We will also impose this scenario for the same situation.

The light intensity with certain frequency $I^{\nu}$ are related to the frequency through the Lorenz invariant as $I^\nu_e/\nu_e^3 = I^\nu_o/\nu_o^3$ \cite{M:2022pex}. Integrating the intensity through the whole frequency range, we obtain the total observed intensity
\begin{equation}
	I_o(r) = \int I^\nu_o(r) d\nu_o = \tilde{g}^4 I_e(r).
\end{equation}

\subsection{Generating shadow image and photon trajectory}

In this subsection, we compare the resulting shadow image of our model with that of the Hayward regular black hole. We assume that light rays can pass through the star's surface without interacting with the medium inside the interior. The differences in the shadow images between our model and the Hayward regular black hole, which contains a horizon, are expected to be significant. In spacetimes with horizon, light rays are absorbed as they reach the horizon, creating a dark region inside the observed photon sphere radius. In contrast, in horizonless spacetimes, all light rays continue to the point of closest approach and are then sent back to infinity. As a result, spacetimes with horizon produce a dark region inside the \textit{observed} photon sphere radius, while horizonless spacetimes can potentially produce multiple \textit{light rings} \cite{Guerrero:2022msp}. We will also compare the resulting image of our star model with a horizonless configuration of Hayward spacetime using the same $x$ value.

\begin{figure}[htbp!]
	\centering
	\includegraphics[width=0.7\textwidth]{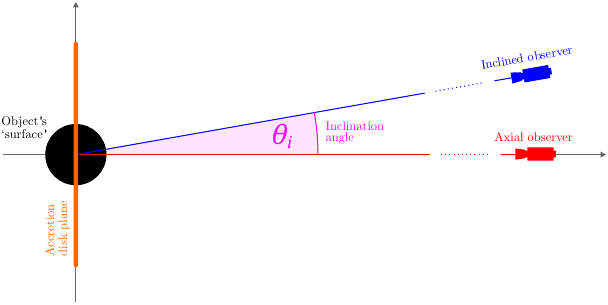}
	\caption{Two-dimensional illustration of axial and inclined observations.}
	\label{fig. incl illustration}
\end{figure}

We analyze the shadow images from various viewing angles. For accretion disk perpendicular to the observer's line of sight, we use the \textit{axial observation} approach. We use the \textit{inclined observation} approach for non-perpendicular axes. Both scenarios are illustrated in Fig.~\ref{fig. incl illustration}. The axial observation helps study the shadow and light ring structure around the object, as its symmetry allows for more efficient image generation with higher resolution. Conversely, the inclined observation is beneficial for producing a more realistic shadow image model by considering the relativistic beaming effect.

\subsubsection{Photon trajectory}

\begin{figure*}[htbp!]
	\centering
	\includegraphics[width=0.8\textwidth]{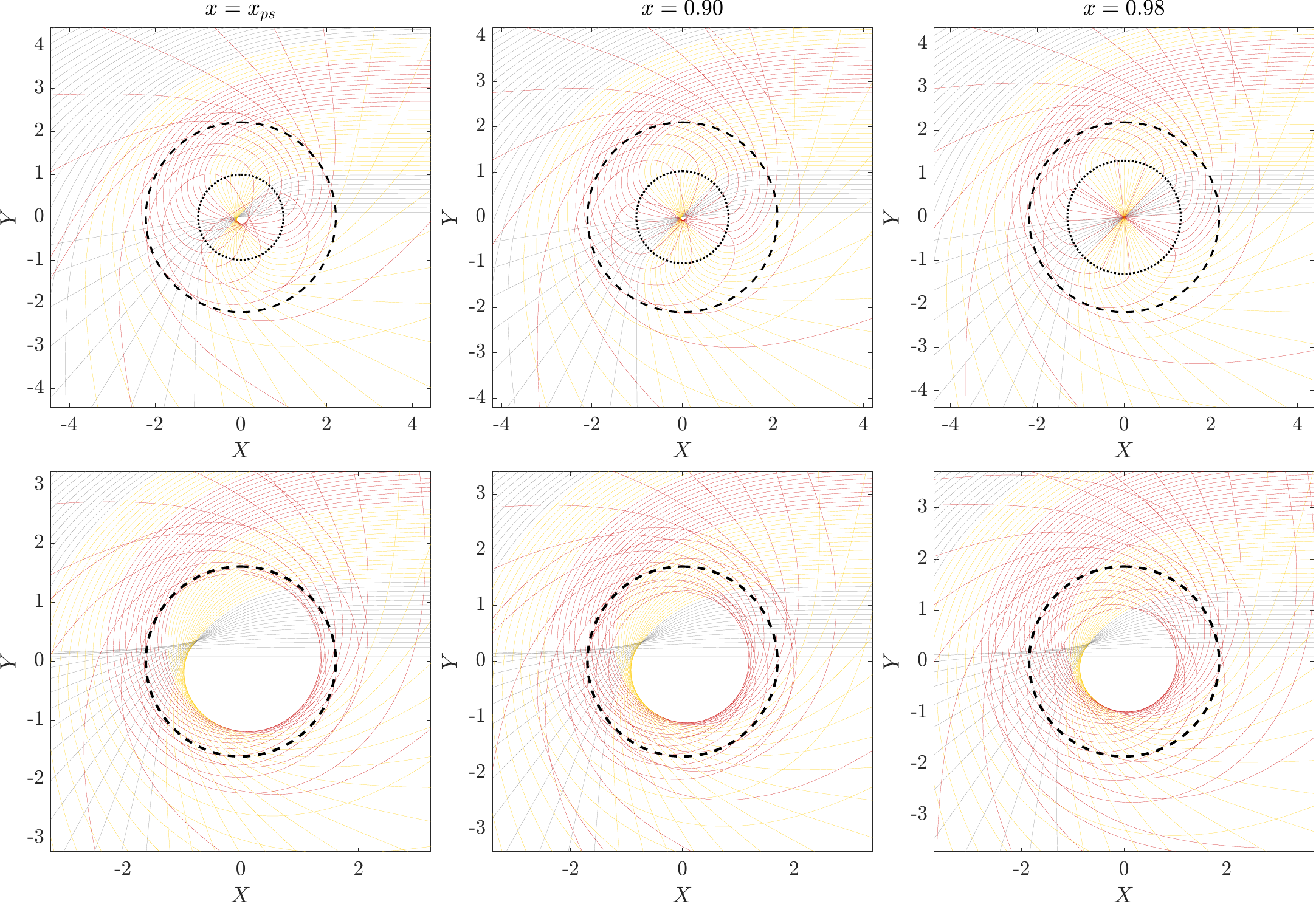}
	\caption{Photon trajectory around our corresponding gravastar model with $g(r) = 1$ (top) and in Hayward spacetime (bottom). Dashed line on the top plot represent the star surface, while on the bottom plot it represent the Schwarzschild radius $R_S=2M_H$. The dotted line on the top plot denotes the outer `crust' where $dp/dr=0$.}
	\label{fig. photon trajectory}
\end{figure*}

\begin{figure}[htbp!]
	\centering
	\includegraphics[width=0.7\textwidth]{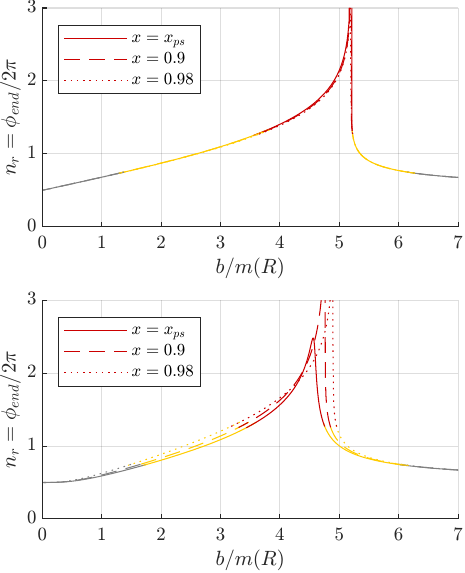}
	\caption{Number of photon orbits $n_r$ for our gravastar model (top) and Hayward spacetime (bottom) as a function of impact parameter $b/M$.}
	\label{fig. endphi impact}
\end{figure}

Before examining the shadow image, it is important to analyze the photon trajectories around the object. We compare the photon trajectories for our gravastar model with those of the horizonless configuration of Hayward-type spacetime. We choose the value of $x$ to be $x=[x_{ps}, 0.9,0.98]$ which corresponds to the total mass of $m_G(R) = [0.7383, 0.7676, 0.8455]$ for our gravastar model and $m_H(R)=M_H=[0.8063, 0.8504, 0.9260]$ for Hayward spacetime. We also analyze the photon's number of orbits defined as
\begin{equation}
	n_r = \frac{\phi_{end}}{2\pi}    
\end{equation}
where $\phi_{end}$ is the final angle of the light ray at numerical infinity. We classify three types of photon trajectories based on the number of orbits (see, e.g., \cite{Zeng:2023fqy,Meng:2023htc}): \textit{direct emmision} ($n_r \leq 3/4$), \textit{lensed ring} ($3/4<n_r<5/4$), and \textit{photon ring} ($n_r \geq 5/4$). The resulting trajectory and the number of orbits for photon impact parameter are given in Figs.~\ref{fig. photon trajectory} and \ref{fig. endphi impact}, respectively.

These results show that the photon trajectories differ significantly between our gravastar model and the horizonless Hayward spacetime. In our gravastar model, photons are strongly deflected toward the center near the crust and follow a straight path through the gravastar's core. In contrast, there is no extreme deflection in horizonless Hayward spacetime, and only photons with low-impact parameters pass near the center. This finding can be easily verified from the photon effective potential shown in Fig.~\ref{fig. photon eff potential}. The photon trajectories in our gravastar model are similar to what Ref.~\cite{Chen:2024ibc} obtained for a semi-classical horizonless ultracompact object, which is expected since it has a similar spacetime structure.

The straight path inside the gravastar's core is expected since the core is de Sitter. Similar straight paths in the de Sitter interior of thin-shell gravastars have been reported in several studies, e.g., \cite{Rosa:2024bqv,Kubo:2016ada}. However, in the original gravastar model, the thin-shell causes light rays to be slightly deflected from the exterior to the interior, with the angle of deflection depending on the thin-shell mass. In our gravastar model, a more massive configuration results in a higher deflection angle at the gravastar crust, directing the photons toward the center.

\subsubsection{Axial observation}
\label{sec. axial observation}

\begin{figure}[htbp!]
	\centering
	\includegraphics[width=.7\textwidth]{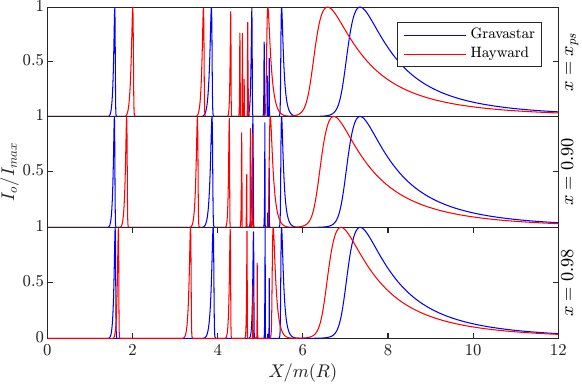}
	\caption{The observed intensity of GLM1 (ISCO) accretion disk model in axial observation for our gravastar model (blue) and Hayward spacetime (red) with arbitrary units normalized by its maximum intensity.}
	\label{fig. inten axial isco}
\end{figure}

\begin{figure*}[htbp!]
	\centering
	\includegraphics[width=.8\textwidth]{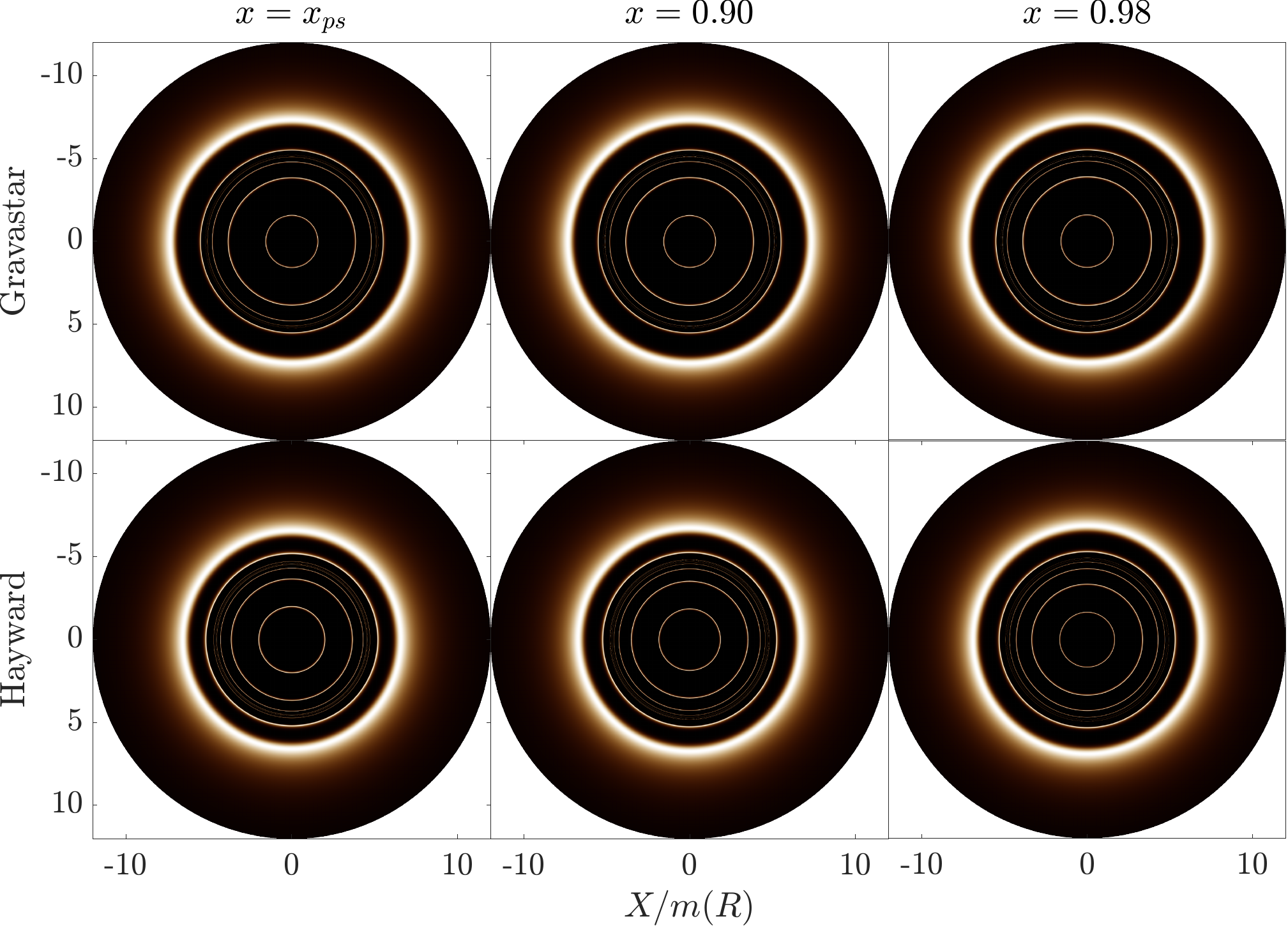}
	\caption{Shadow images of GLM1 (ISCO) accretion disk model in axial observation for our gravastar model (top) and Hayward spacetime (bottom).}
	\label{fig. shadow axial isco}
\end{figure*}

\begin{figure}[htbp!]
	\centering
	\includegraphics[width=.7\textwidth]{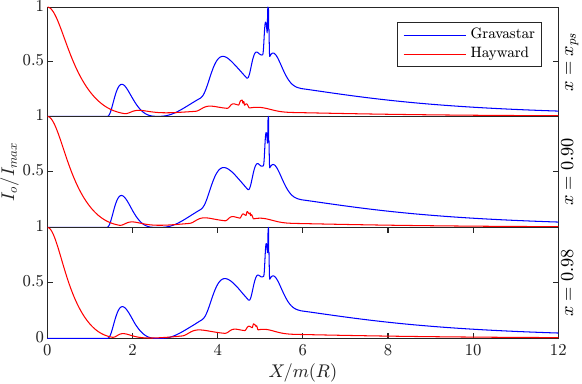}
	\caption{The observed intensity of GLM2 (Center) accretion disk model in axial observation for our gravastar model (blue) and Hayward spacetime (red) with arbitrary units normalized by its maximum intensity.}
	\label{fig. inten axial center}
\end{figure}

\begin{figure*}[htbp!]
	\centering
	\includegraphics[width=.8\textwidth]{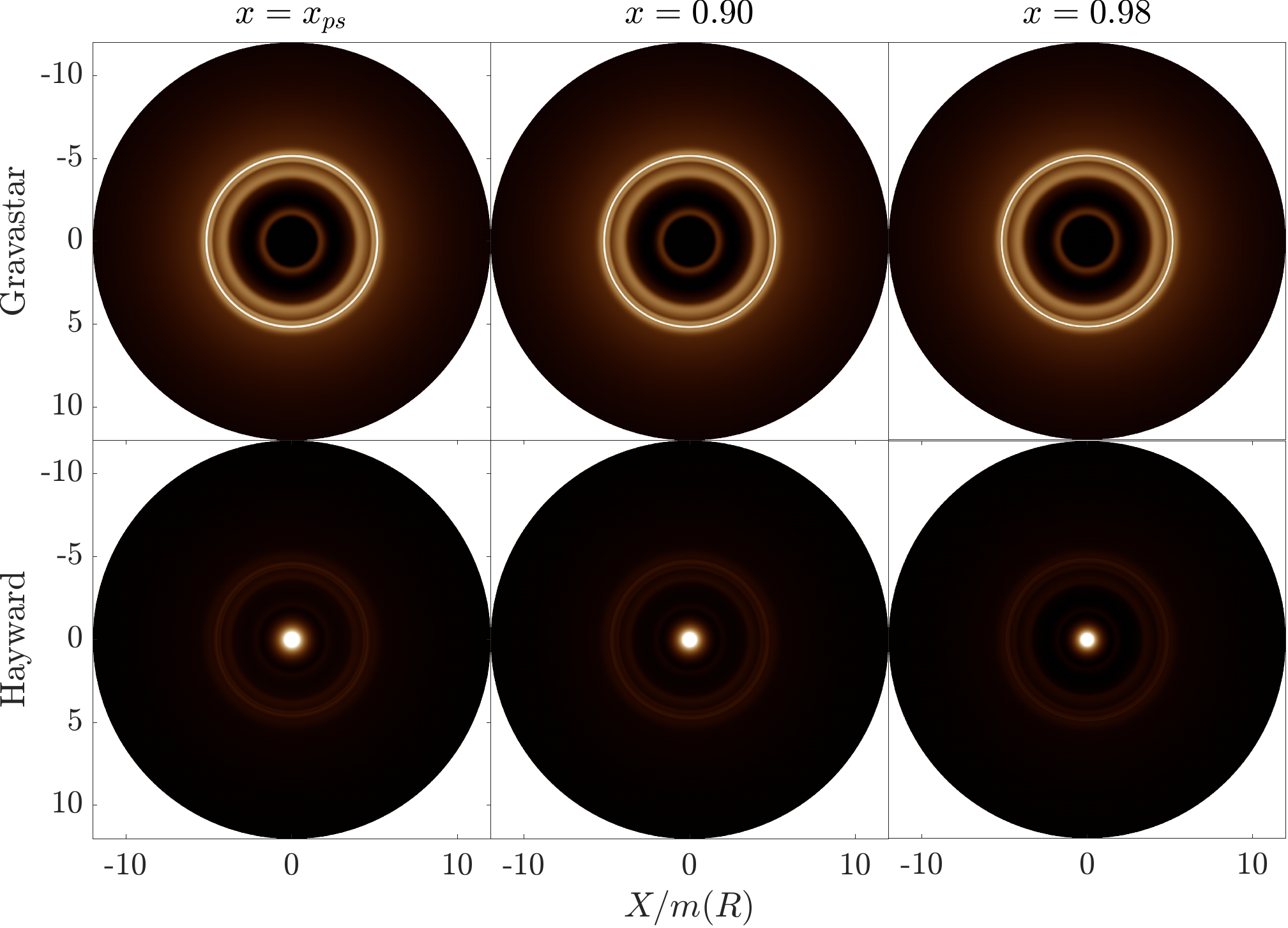}
	\caption{Shadow images of GLM2 (Center) accretion disk model in axial observation for our gravastar model (top) and Hayward spacetime (bottom).}
	\label{fig. shadow axial center}
\end{figure*}

Using spherical symmetry, we generate shadow images for axial observation. This calculation involves evaluating the photon trajectories that cross the middle row of the screen. The recorded intensity in each pixel is then plotted, as shown in Figs.~\ref{fig. inten axial isco} and \ref{fig. inten axial center}. Each resulting intensity is normalized by its maximum value. To obtain the shadow image, we ``rotate" the middle part of the screen, allowing it to sweep across the other areas. The results are shown in Figs.~\ref{fig. shadow axial isco} and \ref{fig. shadow axial center}.

\textit{GLM1 (ISCO) model}\textemdash We observe a similar light ring structure for both models. We classify two types of light rings based on the width of their observed intensity peaks; 
\textit{major} light rings and \textit{minor} light rings. The former has a particular `width' of peak observed intensity, while the latter only peaked at one point. Both models mainly consist of four major light rings and several minor ones between the first and the second outer major light rings. The minor light ring is produced by the exponential growth of the number of orbits near the critical impact parameter (shown by red lines in Figs.~\ref{fig. photon trajectory} and \ref{fig. endphi impact}). It can be seen from Fig.~\ref{fig. inten axial isco} that our gravastar model produces the same relative location of the major light rings for every value of $x$, and it is not the case for Hayward spacetime, where we notice a smaller inner light ring for higher $x$ despite larger peak intensity location near the $R_{ISCO}$.

Comparing our results with thin-shell gravastar images from \cite{Rosa:2024bqv}, we observe significant differences in the structure and behavior of the inner light rings. As demonstrated in the study above, thin-shell gravastars with a photon sphere, i.e., $M/R > 3$, produce three distinct light rings, and their locations depend on compactness and mass distribution. This behavior is not present in our model; instead, we find that for every value of $x$\textemdash which determines the compactness\textemdash as long as a photon sphere is produced ($x > x_{ps}$), the light ring structure remains remarkably consistent.

\textit{GLM2 (Center) model}\textemdash We observe significantly different intensity profiles (Fig.~\ref{fig. inten axial center}) and shadow images (Fig.~\ref{fig. shadow axial center}) produced by the two models. In our gravastar model, a dark patch (zero intensity) appears near the center of the image due to strong redshift effects. This finding is expected since the $e^{\nu}$ component of the metric approaches zero as it nears the center (see Fig.~\ref{fig. metric gconst}). In contrast, this feature is absent in Hayward spacetime because the $e^{\nu}$ metric component approaches unity at the center. Consequently, Hayward spacetime shows peak intensity at its center, while our gravastar model reaches its peak observed intensity at the photon sphere location. This behavior is expected since we use an optically thin accretion disk model, where the intensity accumulates every time a light ray intersects the accretion disk. The images of our anisotropic gravastar also exhibit a structure that differs from, yet bears similarities to, the thin-shell gravastar model, particularly the configuration where the mass is distributed entirely on the thin shell \cite{Rosa:2024bqv}.

\subsubsection{Inclined observation}
\label{sec. inclined observation}

To produce more realistic inclined shadow images, we use a different approach. We employ a complete ray tracing procedure, sending light rays through every pixel on the screen and recording the detected accretion disk intensity. We set the screen resolution to $1000 \times 1000$ pixels to achieve acceptable image quality efficiently. We also incorporate the accretion flow calculation to observe the effects of the relativistic Doppler effect, as discussed in Sec. \ref{sec. Thin accretion disk}. Additionally, we apply a Gaussian filter to the images with a specific standard deviation, $\sigma$, to simulate the optical resolution of a realistic observing instrument like the Event Horizon Telescope (EHT).

We set specific parameter values for our inclined observation images. We select two inclination angles, $\theta_i$, which are $17^\circ$ (the known inclination angle of M87* \cite{Tamburini:2019vrf}) and $60^\circ$. Additionally, we choose $x = 0.9$ for both gravastar and Horizonless Hayward Spacetime model and $x=1.01$ for the Hayward black hole model.

\begin{figure*}[htbp!]
	\centering
	\includegraphics[width=.8\textwidth]{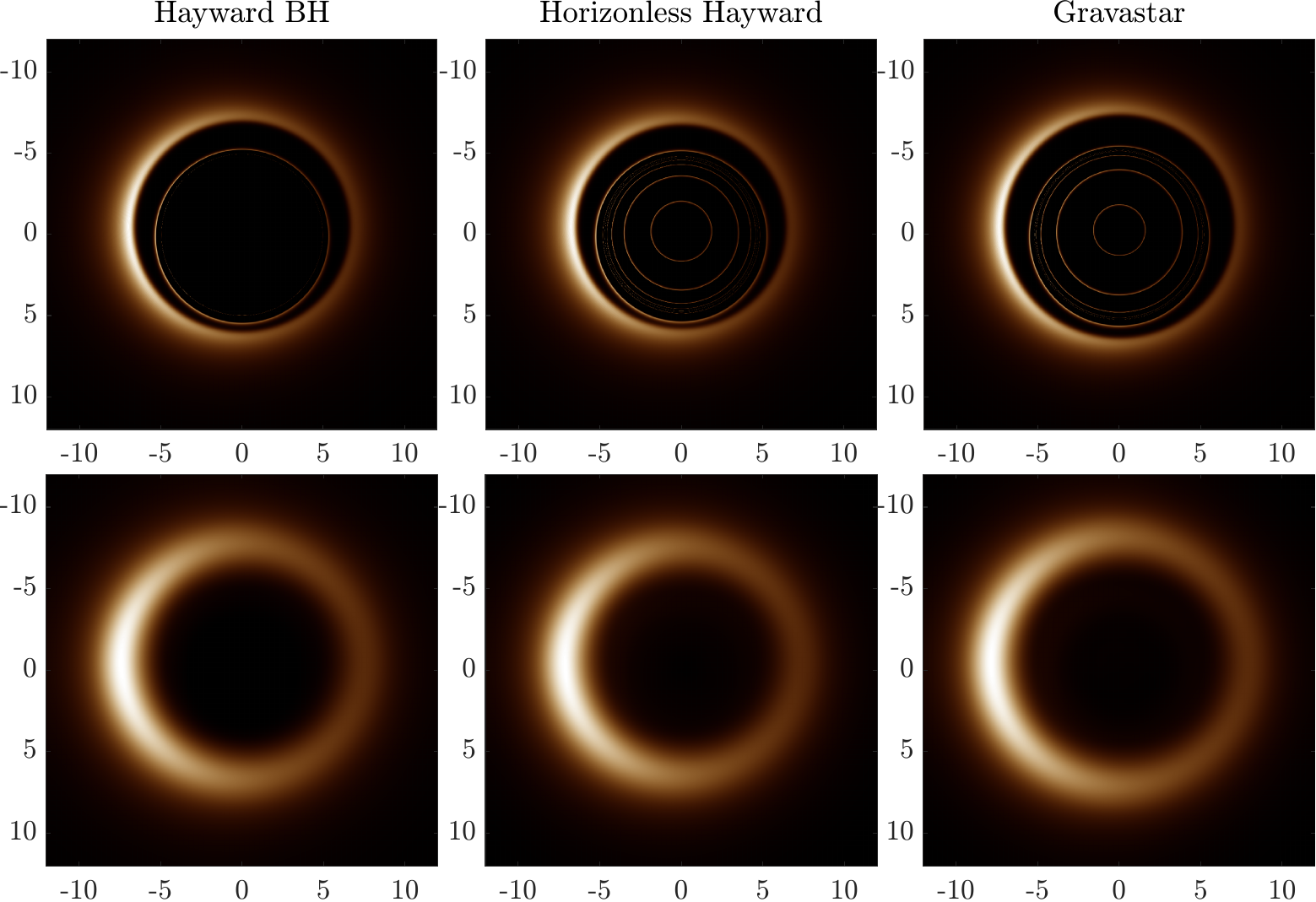}
	\caption{Shadow images of inclined observation with $\theta_i = 17^\circ$. The horizontal and vertical axis are the celestial coordinates normalized by its mass, that is $X/m(R)$ and $Y/m(R)$, respectively.}
	\label{fig. shadow inclined 17deg}
\end{figure*}

\begin{figure}[htb!]
	\centering
	\includegraphics[width=.7\textwidth]{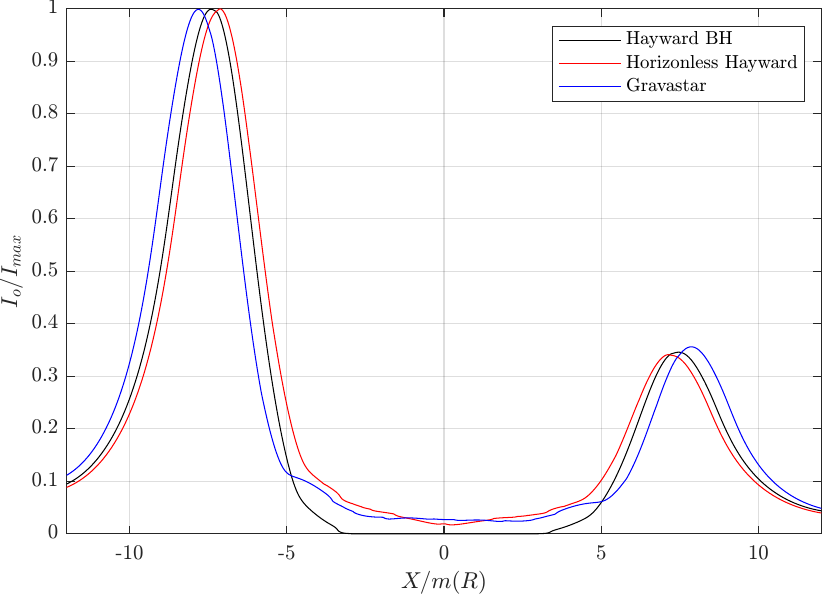}
	\caption{Observed intensities from the center-horizontal cross section of the filtered inclined observation images with  $\theta_i = 17^\circ$.}
	\label{fig. inten inclined 17deg}
\end{figure}

\begin{figure*}[htb!]
	\centering
	\includegraphics[width=.8\textwidth]{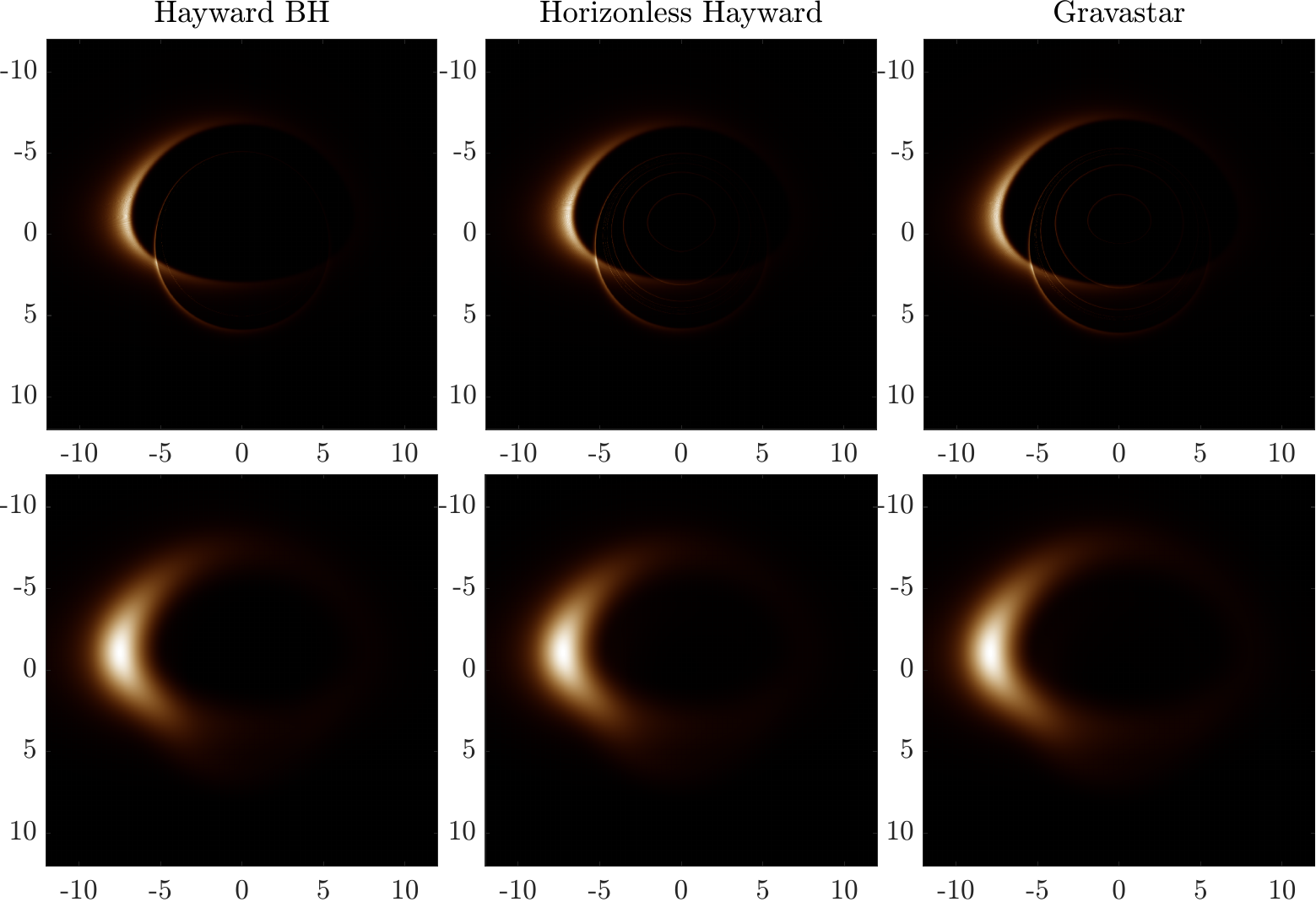}
	\caption{Shadow images of inclined observation with $\theta_i = 60^\circ$.}
	\label{fig. shadow inclined pi3}
\end{figure*}

\begin{figure}[htb!]
	\centering
	\includegraphics[width=.7\textwidth]{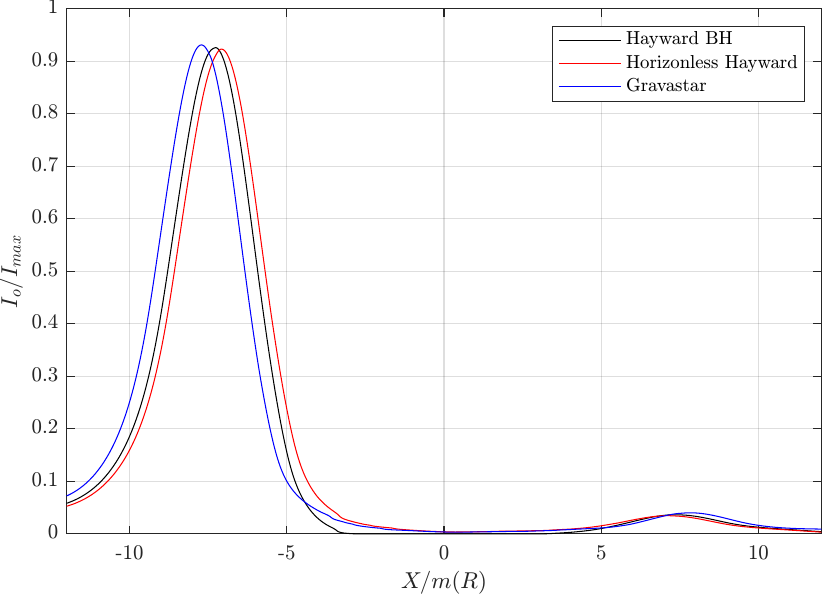}
	\caption{Observed intensities from the center-horizontal cross section of the filtered inclined observation images with  $\theta_i = 60^\circ$.}
	\label{fig. inten inclined pi3}
\end{figure}

After applying the Gaussian filter for the $17^\circ$ inclination angle, as shown in Fig.~\ref{fig. shadow inclined 17deg}, we observe slight differences in the central part of the image. In the black hole configuration, the light ring is fully resolved near the direct emission image, resulting in a completely dark center. In contrast, the horizonless Hayward spacetime and our gravastar model produce a slightly brighter center due to the resolved inner light ring. This central brightness could potentially probe the existence of a horizon, as suggested by \cite{Eichhorn:2022fcl}, with higher-resolution next-generation instruments. However, the differences between our gravastar model and the horizonless Hayward spacetime images are minimal, making them indistinguishable from current and near-future observations. However, as shown in Figs.~\ref{fig. shadow inclined pi3}-\ref{fig. inten inclined pi3}, a larger inclination angle results in a dimmer inner light ring, making it less observable from the black hole configuration.

We also observe brightness asymmetry between the right and left sides of the image caused by the relativistic Doppler effect. This asymmetric feature is also present in the shadow images of M87* \cite{EventHorizonTelescope:2019dse} and Sagittarius A* \cite{EventHorizonTelescope:2022wkp}. From the resulting cross-section intensities of Fig.~\ref{fig. inten inclined 17deg}, we can see that both Hayward spacetimes exhibit slightly greater asymmetry between the peak brightness on the right and left sides compared to the asymmetry produced by the Schwarzschild spacetime in our gravastar model. Assuming the intensity of the accretion disk is axially homogeneous, it is possible to determine the corresponding spacetime from the shadow images by measuring this brightness asymmetry (see, for instance, \cite{Younsi:2021dxe,Guo:2023grt}). As shown in Figs.~\ref{fig. shadow inclined pi3} and \ref{fig. inten inclined pi3}, larger inclination angles increase brightness asymmetry, but the difference in brightness asymmetry between both Hayward spacetimes and our gravastar model becomes indistinguishable. Nevertheless, it should be noted that this difference in asymmetry could be caused by the different spacetime or mass configurations, considering that the same $x$ value does not result in the same total mass for the horizonless Hayward model and our gravastar model.

%=======================================================================
\section{Conclusion}
\label{sec. conclusion}

Over the past decades, extensive studies of regular black holes and horizonless compact objects have led to ongoing questions about the existence of singularities and horizons. Regular black holes possess the intriguing feature of allowing the absence of horizons without violating the {\it Weak Cosmic Censorship Conjecture} (WCCC). In addition to regular black holes, several phenomenological studies of horizonless black hole mimickers have been proposed, including the gravastar. In this study, we have explored the possibility of integrating these two objects into a single framework.

We have constructed an EoS that can describe two objects: regular black holes and anisotropic gravastars. The EoS describes a regular black hole beyond the extremal limit with a de Sitter (dS) EoS $p=-\epsilon$ while fulfilling the requirements of an anisotropic gravastar when the horizon is absent. We also incorporate the need for a smooth cut-off function, which results in a transition region between the black hole configuration and the gravastar configuration. Notably, this EoS in the gravastar configuration incidentally produces a similar spacetime structure\textemdash, particularly in the $e^{\nu}$ component\textemdash to the quantum horizonless compact object (QHCO) described in \cite{Chen:2024ibc}. Consequently, the anisotropic gravastar modeled in this study may intersect with the QHCO.

Regarding the shadow image, by assuming no interaction occurs between the photon and the star interior, we observe both significant and insignificant differences when comparing the horizonless spacetime used in this study, namely the Hayward spacetime, to the anisotropic gravastar. While the light ring structure in the shadow image involving an ISCO accretion disk differs noticeably from that of a regular black hole spacetime with a horizon, it exhibits some similarities between the anisotropic gravastar and the horizonless Hayward spacetime as both configurations feature four major light rings and several minor light rings between the first and second outer major light rings. Significant differences are observed with a center model accretion disk, as the horizonless Hayward spacetime does not produce darkened intensities in the image's central region. Comparing our results to the thin-shell gravastar given by Ref.~\cite{Rosa:2024bqv}, we see the opposite situation as it produces a distinct light ring structure for the ISCO accretion disk while exhibiting similarities with the central model in a specific thin-shell gravastar configuration. Taking the image generation further by incorporating the inclined observation and accretion flow, we found that, at lower resolutions involving Gaussian blur as a natural approach, images of both horizonless spacetimes feature slightly higher intensities in the central part of the image. Additionally, we observe that the brightness asymmetry is somewhat higher in Hayward spacetime than in Schwarzschild spacetime\footnote{It is referred to as Schwarzschild spacetime since the exterior of the anisotropic gravastar\textemdash where the accretion disk is located\textemdash is Schwarzschild}. It should be noted that significant differences in the shadow images may arise if the time delay mechanism due to strong redshifts\textemdash as described in Ref.~\cite{Chen:2024ibc}\textemdash as well as photon interactions with the star's interior, are considered. These factors could result in further darkening of the central images in realistic observational time scale scenarios.

Our research shows that probing horizonless spacetimes between the RBH and anisotropic gravastar model with a photon sphere through optical observation is quite tricky. The most significant differences can be observed when emitting matter is allowed to exist \textit{inside} the object, where the gravitational redshift effects differ considerably between the two spacetimes. If the emitting matter exists only \textit{outside} the objects, it becomes challenging to determine whether the horizonless object corresponds to the anisotropic gravastar or the horizonless RBH configuration, particularly due to the finite resolution of observational instruments. Nonetheless, the inner light ring structure difference between our anisotropic gravastar and the thin-shell gravastar makes it more amenable to detection. However, high-resolution instruments would be necessary to resolve these inner light rings adequately.

Despite the results, several caveats should be noted from this research. We have not considered the object's stability under any perturbation. As shown in Fig.~\ref{fig. photon eff potential}, our model produces stable photon orbits in the interior region, as it was proved in Ref.\cite{Cunha:2017qtt} that horizonless ultracompact, axisymmetric, and stationary spacetimes always possess at least one stable photon orbit. However, it is also known that stable photon orbits can lead to instabilities in several models of ultracompact objects \cite{Cardoso:2014sna,Cunha:2022gde}. We have also not considered the dynamics of the transition mechanism, which could be crucial for explaining the formation of anisotropic gravastars from regular black holes or vice versa. Future research should investigate these issues and examine several other observational signatures, such as quasinormal modes and the thermodynamic properties of the object.

%%%%%%%%%%%%%%%%%%%%%%%%%%%%%%%%%%%%%%%%%%%%%%%%%%%%%%%%%%%%%%%%%%%%%%%%
\begin{acknowledgements} 
AS is funded by PUTI grant 2024-2025 No. NKB-380/UN2.RST/HKP.05.00/2024. 
\end{acknowledgements}

\end{document}